\documentclass[12pt]{article}

\usepackage{graphics}
\usepackage{epsfig}
\usepackage{cite}
\input epsf

\usepackage{amssymb,amsmath,euscript}
\usepackage{indentfirst}
\usepackage{cite}

\title{Isolated prompt photon pair production at hadron colliders with $k_T$-factorization}

\author{A.V.~Lipatov\footnote{lipatov@theory.sinp.msu.ru}}

\begin{document}

\maketitle

\begin{center}

{\it Skobeltsyn Institute of Nuclear Physics,\\ 
Lomonosov Moscow State University,\\ 119991 Moscow, Russia}

\end{center}

\vspace{0.5cm}

\begin{center}

{\bf Abstract }

\end{center}

In the framework of the $k_T$-factorization approach, the isolated prompt photon pair
production in $pp$ and $p \bar p$ collisions at high energies is studied. 
The consideration is based on the quark-antiquark annihilation, quark-gluon scattering 
and gluon-gluon fusion subprocesses, where the non-zero transverse momenta of incoming
partons are taken into account.
The unintegrated quark and gluon densities in a proton are determined using the 
Kimber-Martin-Ryskin prescription. 
The numerical analysis covers the total and differential production cross sections 
and extends to specific angular correlations between the produced prompt photons.
Theoretical uncertainties of our evaluations are studied and
comparison with the NLO pQCD calculations is performed.
The numerical predictions are compared with the recent
experimental data taken by the D$\emptyset$, CDF, CMS and ATLAS collaborations at the Tevatron
and LHC energies.

\vspace{1.0cm}

\noindent
PACS number(s): 12.38.-t, 13.60.Hb, 13.85.Qk

\newpage

\section{Introduction} \indent

The production of prompt photon pairs\footnote{Usually the photons are called ”prompt” if they are
coupled to the interacting quarks.} in hadronic collisions at high energies 
is of great interest at present\cite{1,2,3,4,5}. It provides a major 
background to searches for rare or exotic processes in both the Tevatron and the 
LHC. In particular, it represents a major source of large 
and irreducible background to searches for the Higgs boson in the
low mass range, where its first indications (with mass of about 125--126 GeV) 
have been found very recently.
It is also a significant background to searches for a new heavy resonances\cite{6}, 
cascade decays of a new heavy particles\cite{7} and to searches for
some effects of physics beyond the Standard Model. For example, so called Universal Extra-Dimensions
predict non-resonant diphoton production associated with significant missing 
transverse energy\cite{8,9}. Other extra-dimension models, such as 
Randall-Sundrum model\cite{10}, predict the production of gravitons, which would decay 
into photon pairs with a narrow width.
From another side, studies of prompt diphoton production provide 
a particularly clean test of perturbative Quantum Chromodynamics (pQCD) and
soft-gluon resummation methods implemented in theoretical calculations 
since the produced photons are largely insensitive to the effects of final-state hadronization, 
and their energies and directions can be measured with high precision.
Therefore, it is essential to have an accurate pQCD predictions for 
corresponding cross sections and related kinematical distributions.

The leading contributions to the production of prompt photon pairs at hadron collisions
are the quark-antiquark annihilation $q\bar q \to \gamma \gamma$, gluon-gluon fusion 
$gg \to \gamma \gamma$ and quark-gluon scattering $qg \to \gamma \gamma q$ subprocesses.
Prompt photons may also originate from single or double fragmentation
processes of the partons produced in the hard interaction\cite{11}. 
However, a photon isolation requirement which involved 
in the measurements\cite{1,2,3,4,5} significantly reduces
the rate for these processes\footnote{See discussion in Section~2.}. 
In the framework of standard QCD, theoretical calculations of diphoton
production cross sections have been carried out at next-to-leading order (NLO)\cite{12} and 
next-to-next-to-leading-log approximation (NNLL)\cite{13}.
A fixed-order NLO pQCD calculations\cite{12} implemented by the \textsc{diphox} program 
accounts fragmentation subprocesses, but the gluon-gluon fusion $gg \to \gamma \gamma$
contribution is considered only at LO.
In the NNLL calculations\cite{13} implemented by the \textsc{resbos} program
the effects of initial state soft gluon radiation in
the NLO calculations are analytically resummed to
all orders in the strong coupling constant.
Both these calculations reproduce the main features of recent Tevatron and LHC 
data\cite{1,2,3,4,5}, but none of them describes all aspects of the data.
So, there is significant underestimation\cite{4} of diphoton cross sections measured by the CMS 
collaboration in the regions of phase space where two photons 
have an azimuthal angle difference $\Delta \phi_{\gamma \gamma} \le 2.8$. 
The similar observation was made\cite{5} by the ATLAS collaboration:
more photon pairs are seen in data at low $\Delta \phi_{\gamma \gamma}$ values, 
while the NLO and NNLL pQCD predictions favour a 
larger back-to-back production ($\Delta \phi_{\gamma \gamma} \sim \pi$).
At the Tevatron energies, there is the same problem\cite{1,2,3} common to both NLO\cite{12} and NNLL\cite{13} pQCD 
calculations in the description of events with low diphoton mass $M_{\gamma \gamma}$
or low azimuthal angle distance $\Delta \phi_{\gamma \gamma}$ 
(and also with moderate diphoton transverse momentum $p_T^{\gamma \gamma}$).
Such disagreement between the collinear QCD predictions and the available data indicates\cite{1,2,3,4,5} 
the necessity of including higher order corrections beyond NLO. 

We note, however, that an alternative description of diphoton production 
at high energies can be achieved within the framework of $k_T$-factorization QCD approach\cite{14,15}.
This approach is based on famous Balitsky-Fadin-Kuraev-Lipatov (BFKL)\cite{16} or 
Ciafaloni-Catani-Fiorani-Marchesini (CCFM)\cite{17} equations and 
provides solid theoretical grounds for the effects of initial state gluon radiation and intrinsic
non-zero parton transverse momentum.
A detailed description and discussion of the $k_T$-factorization formalism can be
found, for example, in reviews\cite{18}. Here we would like to only mention 
that the main part of high-order radiative QCD corrections is 
naturally included into the leading-order $k_T$-factorization formalism.
Moreover, the soft gluon resummation formulas implemented in NNLL calculations\cite{13} 
are the result of the approximate treatment of the 
solutions of CCFM equation, as it was shown in\cite{19}.
The $k_T$-factorization approach has been successfully applied recently, in particular, 
to describe the prompt photon and Drell-Yan pair production at HERA, Tevatron and LHC\cite{20,21,22,23,24,25}.

In the present paper we apply the $k_T$-factorization QCD approach to diphoton 
production in $p\bar p$ and $pp$ collisions at high energies.
Our main goal is to give a systematic analysis of available
Tevatron data taken by the D$\emptyset$ and CDF collaborations\cite{1,2,3} and 
recent LHC measurements\cite{4,5} performed by the CMS and ATLAS 
collaborations.
The consideration is based on the off-shell (depending on the non-zero transverse momenta $k_T$ of 
incoming quarks and gluons) production amplitudes of quark-antiquark annihilation $q\bar q \to \gamma \gamma$, 
quark-gluon scattering $qg \to \gamma \gamma q$ and gluon-gluon 
fusion $gg \to \gamma \gamma$ subprocesses\footnote{In the case of gluon-gluon fusion
we will neglect the transverse momenta of incoming gluons in the corresponding production amplitude 
but keep true off-shell subprocess kinematics. See Section~2 for more details.}.
The unintegrated ($k_T$-dependent) quark and gluon densities in a proton
are defined by using the Kimber-Martin-Ryskin (KMR) prescription\cite{26}.
This approach is the simple formalism to construct the unintegrated parton 
distributions from the known conventional ones.
We calculate total and differential diphoton production cross sections
and estimate the theoretical uncertainties of our predictions.
Special attention is put to the specific
kinematic properties of the produced photon pair which are 
strongly related to the non-zero transverse momenta of initial partons.
Such investigations in the framework of $k_T$-factorization approach 
are performed for the first time.

The outline of our paper is following. In Section~2 we 
recall shortly the basic formulas of the $k_T$-factorization approach with a brief 
review of calculation steps. In Section~3 we present the numerical results
of our calculations and a discussion. Section~4 contains our conclusions.

\section{Theoretical framework} \indent

As it was mentioned above, the leading contributions to the diphoton
production at high energies are the quark-antiquark annihilation, 
quark-gluon scattering and gluon-gluon fusion subprocesses:
\begin{gather}
q(k_1) + \bar q(k_2) \to \gamma(k_3) + \gamma(k_4),\\
q(k_1) + g(k_2)\to \gamma(k_3) + \gamma(k_4) + q(k_5),\\
g(k_1) + g(k_2) \to \gamma(k_3) + \gamma(k_4),
\end{gather}

\noindent
where the four-momenta of all particles are given in parentheses. 
Although last two of them are higher-order processes, their contributions are 
quantitatively comparable to those from quark-antiquark annihilation in the diphoton
invariant mass range of interest, due to the significant gluon luminosity in this kinematical 
region.

Let us start from the kinematics.
In the center-of-mass frame of colliding particles we can write:
\begin{equation}
p^{(1)} = {\sqrt s\over 2} (1,0,0,1),\quad p^{(2)} = {\sqrt s\over 2} (1,0,0,-1),
\end{equation}

\noindent
where $p^{(1)}$ and $p^{(2)}$ are the four-momenta of colliding protons, 
$\sqrt s$ is the total energy of the process under consideration and we neglect the 
masses of protons. The initial parton four-momenta in the high energy limit can be
written as follows:
\begin{equation}
k_1 = x_1 p^{(1)} + k_{1T}, \quad k_2 = x_2 p^{(2)} + k_{2T},
\end{equation}

\noindent 
where $k_{1T}$ and $k_{2T}$ are their transverse four-momenta. 
It is important that ${\mathbf k}_{1T}^2 = - k_{1T}^2 \neq 0$, 
${\mathbf k}_{2T}^2 = - k_{2T}^2 \neq 0$. From the conservation laws we can easily obtain 
the following relations for subprocesses~(1) and~(3):
\begin{equation}
{\mathbf k}_{1T} + {\mathbf k}_{2T} = {\mathbf k}_{3T} + {\mathbf k}_{4T},
\end{equation}
\begin{equation}
x_1 \sqrt s = |{\mathbf k}_{3T}| e^{y_3} + |{\mathbf k}_{4T}| e^{y_4}, 
\end{equation}
\begin{equation}
x_2 \sqrt s = |{\mathbf k}_{3T}| e^{-y_3} + |{\mathbf k}_{4T}| e^{-y_4},
\end{equation}

\noindent 
and the similar ones for subprocess~(2):
\begin{equation}
{\mathbf k}_{1T} + {\mathbf k}_{2T} = {\mathbf k}_{3T} + {\mathbf k}_{4T} + {\mathbf k}_{5T}, 
\end{equation}
\begin{equation}
x_1 \sqrt s = |{\mathbf k}_{3T}| e^{y_3} + |{\mathbf k}_{4T}| e^{y_4} + m_{5T} e^{y_5}, 
\end{equation}
\begin{equation}
x_2 \sqrt s = |{\mathbf k}_{3T}| e^{-y_3} + |{\mathbf k}_{4T}| e^{-y_4} + m_{5T} e^{-y_5},
\end{equation}

\noindent 
where $k_{3T}$, $k_{4T}$ and $k_{5T}$ are the transverse four-momenta 
of corresponding particles, $y_3$, $y_4$ and $y_5$ are their center-of-mass rapidities
and $m_{5T}$ is the transverse mass of produced quark (or antiquark) having mass $m$.

The off-shell partonic amplitudes of~(1) and~(2) can be written as follows:
\begin{equation}
  \displaystyle  {\cal M}(q^*\bar q^*\to \gamma \gamma) = e^2 e_q^2 \, \epsilon_\mu(k_3) \epsilon_\nu(k_4) \, \times \atop 
  \displaystyle  { \times \bar u(k_2) \left[\gamma^\nu {\hat k_1 - \hat k_3 + m \over (k_1 - k_3)^2 - m^2} \gamma^\mu + 
  \displaystyle  \gamma^\mu {\hat k_1 - \hat k_4 + m \over (k_1 - k_4)^2 - m^2} \gamma^\nu \right] u(k_1) },
\end{equation}
\begin{equation}
  {\cal M}(q^*g^*\to \gamma \gamma q) =  e^2 e_q^2 \, g t^a 
  \epsilon_\lambda(k_2) \epsilon_\mu(k_3) \epsilon_\nu(k_4) \sum_{i = 1}^6 {\cal F}_i^{\mu \nu \lambda},
\end{equation}
\begin{equation}
  {\cal F}_1^{\mu \nu \lambda} = \bar u(k_5) \gamma^\nu {\hat k_4 + \hat k_5 + m\over (k_4 + k_5)^2 - m^2} 
  \gamma^\mu {\hat k_1 + \hat k_2 + m\over (k_1 + k_2)^2 - m^2} \gamma^\lambda u(k_1),
\end{equation}
\begin{equation}
  {\cal F}_2^{\mu \nu \lambda} = \bar u(k_5) \gamma^\mu {\hat k_3 + \hat k_5 + m\over (k_3 + k_5)^2 - m^2} 
  \gamma^\nu {\hat k_1 + \hat k_2 + m\over (k_1 + k_2)^2 - m^2} \gamma^\lambda u(k_1),
\end{equation}
\begin{equation}
  {\cal F}_3^{\mu \nu \lambda} = \bar u(k_5) \gamma^\nu {\hat k_4 + \hat k_5 + m\over (k_4 + k_5)^2 - m^2} 
  \gamma^\lambda {\hat k_1 - \hat k_3 + m\over (k_1 - k_3)^2 - m^2} \gamma^\mu u(k_1),
\end{equation}
\begin{equation}
  {\cal F}_4^{\mu \nu \lambda} = \bar u(k_5) \gamma^\lambda {\hat k_5 - \hat k_2 + m\over (k_5 - k_2)^2 - m^2} 
  \gamma^\mu {\hat k_1 - \hat k_4 + m\over (k_1 - k_4)^2 - m^2} \gamma^\nu u(k_1),
\end{equation}
\begin{equation}
  {\cal F}_5^{\mu \nu \lambda} = \bar u(k_5) \gamma^\mu {\hat k_3 + \hat k_5 + m\over (k_3 + k_5)^2 - m^2} 
  \gamma^\lambda {\hat k_1 - \hat k_4 + m\over (k_1 - k_4)^2 - m^2} \gamma^\nu u(k_1),
\end{equation}
\begin{equation}
  {\cal F}_6^{\mu \nu \lambda} = \bar u(k_5) \gamma^\lambda {\hat k_5 - \hat k_2 + m\over (k_5 - k_2)^2 - m^2} 
  \gamma^\nu {\hat k_1 - \hat k_3 + m\over (k_1 - k_3)^2 - m^2} \gamma^\mu u(k_1),
\end{equation}

\noindent
where $e$ and $e_q$ are the electron and quark (fractional) electric charges, $g$ is the
strong charge, $\epsilon(k_2)$, $\epsilon(k_3)$ and $\epsilon(k_4)$ are the polarization
four-vectors of corresponding particles and $a$ is the eight-fold color index.
When we calculate the matrix elements squared, the summation on the polarizations 
of produced photons is carried out by usual covariant formula:
\begin{equation}
  \sum \epsilon^\mu(k_i) \epsilon^{* \nu}(k_i) = - g^{\mu \nu},
\end{equation}

\noindent
where $i = 3, 4$. In the case of initial off-shell gluon we apply the BFKL prescription\cite{14}:
\begin{equation}
  \sum \epsilon^\mu(k_2) \epsilon^{* \nu}(k_2) = {k_{2T}^\mu k_{2T}^\nu\over {\mathbf k}_{2T}^2},
\end{equation}

\noindent
This formula converges to the usual one (20) after azimuthal angle averaging 
in the $|{\mathbf k}_{2T}| \to 0$ limit. 
Since we do not neglect the transverse momentum of incoming quarks,
the standard on-shell quark spin density matrix has to be replaced by a more complicated expression.
To evaluate it we will follow an approximation proposed in\cite{20}. We "extend" the original 
diagram and consider the off-shell quark line as internal
line in the extended diagram. The ”extended” process looks like follows: the initial on-shell
quark with four-momentum $p$ and mass $m$ radiates a quantum (say, photon or gluon) and
becomes an off-shell quark with four-momentum $k$. So, for the extended diagram squared
we can write:
\begin{equation}
  |{\cal M}|^2 \sim {\rm tr} \left[ \bar {\cal T}^\mu {\hat k + m\over k^2 - m^2} \gamma^\nu\, u(p) \bar u(p) \, 
  \gamma_\nu {\hat k + m\over k^2 - m^2} {\cal T}_\mu \right],
\end{equation}

\noindent 
where $\cal T$ is the rest of the original matrix element.
The expression presented between $\bar {\cal T}^\mu$ and ${\cal T}_\mu$ now plays the role 
of the off-shell quark spin density matrix. Using the standard on-shell condition 
$\sum u(p)\bar u(p) = \hat p + m$ and performing the Dirac algebra one obtains
in the massless limit $m \to 0$:
\begin{equation}
  |{\cal M}|^2 \sim {2\over (k^2)^2} {\rm tr} \left[ \bar {\cal T}^\mu \left( k^2 \hat p - 
  2 (p \cdot k) \hat k\right) {\cal T}_\mu \right].
\end{equation}

\noindent 
Now, using the Sudakov decomposition $k = x p + k_T$ (where $k_T$ is the 
off-shell quark non-zero transverse four-momentum, $k^2 = k_T^2 = - {\mathbf k}_T^2$) 
and neglecting the second term in the
parentheses in~(23) in the small-$x$ limit, we easily obtain:
\begin{equation}
  |{\cal M}|^2 \sim {2\over x k^2} {\rm tr} \left[ \bar {\cal T}^\mu x \hat p {\cal T}_\mu \right].
\end{equation}

\noindent 
Essentially, we have neglected here the negative light-cone momentum fraction of the incoming quark. 
The properly normalized off-shell spin density matrix is given by $x \hat p$, while
the factor $2/x k^2$ has to be attributed to the quark distribution function (determining its
leading behavior). With this normalization, we successfully recover the on-shell collinear 
limit when $k$ is collinear with $p$. 
Further calculations are straighforward and in other respects follow the standard QCD
Feynman rules. We only mention that the method of orthogonal amplitudes\cite{27} has been
applied to avoid the long output. The evaluation of traces in (12) --- (19) was 
done using the algebraic manipulation system \textsc{form}\cite{28}.

The matrix element squared of gluon-gluon fusion subprocess~(3) 
was calculated a long time ago in the on-shell 
limit $|{\mathbf k}_{1T}| \to 0$, $|{\mathbf k}_{2T}| \to 0$.
The simple analytical expression can be found, 
for example, in\cite{29}. In our phenomenological study,
we apply it with one remark: numerically, we keep the exact off-shell kinematics given by 
(5) --- (8).

In according to the $k_T$-factorization theorem, 
to calculate the cross section of diphoton production
one should convolute off-shell partonic cross sections
with the relevant unintegrated quark and/or gluon distributions in a proton:
\begin{equation}
  \sigma = \sum_{i,j = q,\,g} \int {\hat \sigma}_{ij}^*(x_1, x_2, {\mathbf k}_{1T}^2, {\mathbf k}_{2T}^2) \, f_i(x_1,{\mathbf k}_{1T}^2,\mu^2) f_j(x_2,{\mathbf k}_{2T}^2,\mu^2) \, dx_1 dx_2 \, d{\mathbf k}_{1T}^2 d{\mathbf k}_{2T}^2, 
\end{equation}

\noindent
where ${\hat \sigma}_{ij}^*(x_1, x_2, {\mathbf k}_{1T}^2, {\mathbf k}_{2T}^2)$
is the off-shell partonic cross section and $f_i(x,{\mathbf k}_{T}^2,\mu^2)$ is the
unintegrated parton densities in a proton. The contributions from 
quark-antiquark annihilation, quark-gluon scattering and gluon-gluon fusion
can be easily rewritten as follows:
\begin{equation}
  \displaystyle \sigma = \sum_{q} \int {1\over 16\pi (x_1 x_2 s)^2 } |\bar {\cal M}(q^*\bar q^* \to \gamma \gamma)|^2 \times \atop
  \displaystyle  \times f_q(x_1,{\mathbf k}_{1T}^2,\mu^2) f_q(x_2,{\mathbf k}_{2T}^2,\mu^2) d{\mathbf k}_{1T}^2 
  d{\mathbf k}_{2T}^2 d{\mathbf k}_{3T}^2 dy_3 dy_4 {d\phi_1 \over 2\pi} {d\phi_2 \over 2\pi},
\end{equation}
\begin{equation}
  \displaystyle \sigma = \sum_{q} \int {1\over 256\pi^3 (x_1 x_2 s)^2 } |\bar {\cal M}(q^* g^* \to \gamma \gamma q)|^2 \times \atop
  \displaystyle  \times f_q(x_1,{\mathbf k}_{1T}^2,\mu^2) f_g(x_2,{\mathbf k}_{2T}^2,\mu^2) d{\mathbf k}_{1T}^2 d{\mathbf k}_{2T}^2 d{\mathbf k}_{3T}^2 d{\mathbf k}_{4T}^2 dy_3 dy_4 dy_5 {d\phi_1 \over 2\pi} {d\phi_2 \over 2\pi} {d\phi_3 \over 2\pi} {d\phi_4 \over 2\pi},
\end{equation}
\begin{equation}
  \displaystyle \sigma = \int {1\over 16\pi (x_1 x_2 s)^2 } |\bar {\cal M}(g g \to \gamma \gamma)|^2 \times \atop
  \displaystyle  \times f_g(x_1,{\mathbf k}_{1T}^2,\mu^2) f_g(x_2,{\mathbf k}_{2T}^2,\mu^2) d{\mathbf k}_{1T}^2 
  d{\mathbf k}_{2T}^2 d{\mathbf k}_{3T}^2 dy_3 dy_4 {d\phi_1 \over 2\pi} {d\phi_2 \over 2\pi},
\end{equation}

\noindent
where $\phi_1$, $\phi_2$, $\phi_3$ and $\phi_4$ are the azimuthal angles of initial partons
and produced photons, respectively. If we average these expressions over $\phi_1$ and $\phi_2$ and 
take the limit $|{\mathbf k}_{1T}| \to 0$ and  $|{\mathbf k}_{2T}| \to 0$, then we recover the 
corresponding formulas in the collinear QCD factorization.

Concerning the unintegrated quark and gluon densities in 
a proton, we apply the Kimber-Martin-Ryskin (KMR) approach\cite{26} to calculate them. The KMR approach is the formalism
to construct the unintegrated parton distributions from the known conventional ones. 
In this approximation the unintegrated quark and gluon distributions are given by
\begin{equation}
  \displaystyle f_q(x,{\mathbf k}_T^2,\mu^2) = T_q({\mathbf k}_T^2,\mu^2) {\alpha_s({\mathbf k}_T^2)\over 2\pi} \times \atop {
  \displaystyle \times \int\limits_x^1 dz \left[P_{qq}(z) {x\over z} q\left({x\over z},{\mathbf k}_T^2\right) \Theta\left(\Delta - z\right) + P_{qg}(z) {x\over z} g\left({x\over z},{\mathbf k}_T^2\right) \right],}
\end{equation}
\begin{equation}
  \displaystyle f_g(x,{\mathbf k}_T^2,\mu^2) = T_g({\mathbf k}_T^2,\mu^2) {\alpha_s({\mathbf k}_T^2)\over 2\pi} \times \atop {
  \displaystyle \times \int\limits_x^1 dz \left[\sum_q P_{gq}(z) {x\over z} q\left({x\over z},{\mathbf k}_T^2\right) + P_{gg}(z) {x\over z} g\left({x\over z},{\mathbf k}_T^2\right)\Theta\left(\Delta - z\right) \right],} 
\end{equation}

\noindent
where $P_{ab}(z)$ are the usual unregulated LO DGLAP splitting 
functions. The theta functions which appears
in~(29) and~(30) imply the angular-ordering constraint $\Delta = \mu/(\mu + |{\mathbf k}_T|)$ 
specifically to the last evolution step to regulate the soft gluon
singularities. 
Numerically, for the input we have used leading-order parton densities $xq(x,\mu^2)$ and 
$xg(x,\mu^2)$ from MSTW'2008 set\cite{30}.

We note that perturbation theory becomes nonapplicable when the wavelength of the emitted photon
(in the emitting quark rest frame) becomes larger that the typical hadronic scale 
${\cal O}$(1~GeV$^{-1}$). 
Then the nonperturbative effects of hadronization or fragmentation must be taken
into account. Accordingly, the calculated cross section can be split into two pieses
\begin{equation}
  \sigma = \sigma_{\rm dir}(\mu^2) + \sigma_{\rm fragm}(\mu^2),
\end{equation}

\noindent
with $\sigma_{\rm dir}(\mu^2)$ representing the perturbative contribution and 
$\sigma_{\rm fragm}(\mu^2)$ the fragmentation contribution.
In the fragmentation processes photons are produced through
the fragmentation of a parton (produced in a hard subprocess) 
into a single photon carrying a large fraction $z$ of parent parton
momentum. These processes are described in terms of quark-to-photon $D_{q\to \gamma}(z,\mu^2)$ and 
gluon-to-photon $D_{g\to \gamma}(z,\mu^2)$ fragmentation functions\cite{31}.
In our calculations for quark-gluon scattering~(2) we choose the fragmentation scale $\mu^2$ to be the
invariant mass of the quark + photon subsystem, and restrict $\sigma_{\rm dir}(\mu^2)$
to $\mu \ge M \simeq 1$~GeV. Under this condition, the contribution $\sigma_{\rm dir}(\mu^2)$ 
is free from divergences (so the mass of the light quark $m$ can be safely sent to zero)
and we checked that the sensitivity of
our results to the choice of $M$ is reasonably soft. As far as
the fragmentation contribution $\sigma_{\rm fragm}(\mu^2)$ is concerned, its size is dramatically reduced by the photon
isolation cuts implemented in the experimental analysis\cite{1,2,3,4,5}. 
The isolation condition required that the hadronic transverse energy 
$E_T^{\rm had}$, deposited inside a cone with aperture $R$ centered around
the photon direction in the pseudo-rapidity $\eta$ and azimuthal angle $\phi$ plane, 
is smaller than some value $E_T^{\rm max}$. In the recent measurements\cite{1,2,3,4,5}
performed at the Tevatron and LHC, these parameters were taken $R \sim 0.4$ and
$E_T^{\rm max} \sim 1 - 2$~ GeV.
According to the estimates\cite{3}, the contribution
from $\sigma_{\rm fragm}(\mu^2)$ amounts to about 10 --- 15\% of 
the visible cross section and therefore is neglected in our consideration.

The multidimensional integrations in~(26) --- (28) have been performed
by the means of Monte Carlo technique, using the routine \textsc{vegas}\cite{32}.
The full C$++$ code is available from the author on request.

\section{Numerical results} \indent

We now are in a position to present our numerical results. First we describe our
theoretical input and the kinematic conditions. After we fixed the unintegrated
gluon distributions, the cross sections (26) --- (28) depend on
the renormalization and factorization scales $\mu_R$ and $\mu_F$. 
Numerically, we set them to be equal to $\mu_R^2 = \mu_F^2 = (\xi M_{\gamma \gamma})^2$. 
In order to estimate the scale 
uncertainties of our calculations
we vary the parameter $\xi$ between 1/2 and 2 about the default value $\xi = 1$.
Since the expression for the off-shell quark spin density matrix has been derived
in the massless approximation, numerically we neglect the quark masses. 
We use the LO formula for the strong 
coupling constant $\alpha_s(\mu^2)$ with $n_f = 4$ 
active quark flavors at $\Lambda_{\rm QCD} = 200$~MeV, so that $\alpha_s(M_Z^2) = 0.1232$.

Experimental data for prompt photon pair production in $p\bar p$ collisions at the Tevatron come
from both the CDF\cite{1,2} and D$\emptyset$\cite{3} collaborations. Several differential 
cross section have been determined: as a function of the diphoton 
invariant mass $M_{\gamma \gamma}$, the transverse momentum and rapidity of photon pair 
$p_T^{\gamma \gamma}$ and $y_{\gamma \gamma}$, the azimuthal angle difference between the 
produced photons $\Delta \phi_{\gamma \gamma}$ and the cosine of
the polar scattering angle $\theta^*$ of leading photon in the Collins-Soper frame.
Recent D$\emptyset$ data\cite{3} refer to the kinematic region defined by
$|\eta^{\gamma 1}| < 0.9$, $|\eta^{\gamma 2}| < 0.9$, $p_T^{\gamma 1} > 21$~GeV and 
$p_T^{\gamma 2} > 20$~GeV with the total energy $\sqrt s = 1960$~GeV, and an additional
cut $M_{\gamma \gamma} > p_T^{\gamma \gamma}$ was applied. 
The measurements of double differential cross sections 
$d\sigma/dM_{\gamma \gamma} dp_T^{\gamma \gamma}$, $d\sigma/dM_{\gamma \gamma} d\Delta \phi_{\gamma \gamma}$ and 
$d\sigma/dM_{\gamma \gamma} d|\cos \theta^*|$ for three subdivisions of
$M_{\gamma \gamma}$ range (namely $30 < M_{\gamma \gamma} < 50$~GeV, 
$50 < M_{\gamma \gamma} < 80$~GeV and $80 < M_{\gamma \gamma} < 350$~GeV)
have been presented also.
The CDF data\cite{2} refer to the kinematic region defined
by $|y^{\gamma 1}| < 1$, $|y^{\gamma 2}| < 1$, $p_T^{\gamma 1} > 17$~GeV and 
$p_T^{\gamma 2} > 15$~GeV. In this analysis, the diphoton 
cross section as a function of variable $z = p_T^{\gamma 2}/p_T^{\gamma 1}$ , 
the ratio of sub-leading to leading photon transverse momentum ($0 < z < 1$), 
has been measured additionally. Also two specific kinematic cases have been examined separately:
differential cross sections for $M_{\gamma \gamma} > p_T^{\gamma \gamma}$ and
$M_{\gamma \gamma} < p_T^{\gamma \gamma}$. Previous CDF data\cite{1}
refer to the kinematic region defined by $|\eta^{\gamma 1}| < 0.9$, $|\eta^{\gamma 2}| < 0.9$, 
$p_T^{\gamma 1} > 14$~GeV, $p_T^{\gamma 2} > 13$~GeV and $\sqrt s = 1960$~GeV. 

First experimental data for prompt diphoton production in $pp$ collisions at the LHC 
come from the CMS\cite{4} and ATLAS\cite{5} collaborations. 
The CMS data\cite{4} have been obtaned in the kinematic region defined by
$|\eta^{\gamma 1}| < 2.5$, $|\eta^{\gamma 2}| < 2.5$ (except pseudo-rapidity
region $1.44 < \eta < 1.57$ for both photons), $p_T^{\gamma 1} > 23$~GeV and 
$p_T^{\gamma 2} > 20$~GeV with the total energy $\sqrt s = 7$~TeV. The measurements
for more tight central region with $|\eta^{\gamma 1}| < 1.44$ and $|\eta^{\gamma 2}| < 1.44$
have been presented additionally. The ATLAS data\cite{5} refers to the kinematic region
defined by $|\eta^{\gamma 1}| < 2.37$, $|\eta^{\gamma 2}| < 2.37$ (with the exclusion 
of the region $1.37 < \eta < 1.52$ for both photons), $p_T^{\gamma 1} > 16$~GeV and 
$p_T^{\gamma 2} > 16$~GeV with the with the same total energy.

\begin{table}
\begin{center}
\begin{tabular}{|l|c|}
\hline
   & \\
  Source & $\sigma(p\bar p \to \gamma \gamma + X)$ [pb] \\
   & \\
\hline
   & \\
   $k_T$-factorization (KMR) & $8.87^{+1.36}_{-0.98}$ (scales)\\
   & \\
   NLO pQCD\cite{12} (\textsc{diphox}) & $10.58\pm 0.55$\\
   & \\
   NNLL pQCD\cite{13} (\textsc{resbos}) & $11.31\pm 2.45$\\
   & \\
   CDF data\cite{2} & $12.47 \pm 0.21$ (stat.) $ \pm 3.74$ (syst.)\\
   & \\
\hline
\end{tabular}
\end{center}
\caption{The total cross section of diphoton production in $p\bar p$ collisions 
calculated in the kinematical region defined by $|y^{\gamma 1}| < 1$, 
$|y^{\gamma 2}| < 1$, $p_T^{\gamma 1} > 17$~GeV and 
$p_T^{\gamma 2} > 15$~GeV at $\sqrt s = 1960$~GeV.}
\end{table}

The results of our calculations are shown in Figs.~1 --- 10 in 
comparison with the Tevatron data\cite{1,2,3} and in Figs.~11 --- 15 
in comparison with the LHC data\cite{4,5}. 
The solid histograms are obtained by fixing both the
factorization and renormalization scales at the default value,
whereas the upper and lower dashed histograms correspond to the scale variation as it
was described above (in left panels). Also, in Figs.~1 --- 6 we plot for comparison 
the NNLL pQCD predictions\cite{13} (as given by \textsc{resbos} program)
listed in\cite{1,3}. The relative contributions to the diphoton cross sections
from quark-antiquark annihilation, quark-gluon scattering and gluon-gluon fusion subprocesses
are shown separately by the dash-dotted, dotted and dashed histograms, respectively,
in right panels. 

We note that different kinematic variables under consideration probe different aspects 
of the diphoton production mechanism. For instance,
the $M_{\gamma \gamma}$ spectrum is particularly sensitive 
to potential contributions from physics beyond the SM.
The $\cos \theta^*$ distribution probes the 
angular momentum of the final state, which should
be different for QCD-mediated production as compared,
for example, to the decay of a scalar Higgs boson $H \to \gamma \gamma$ (see \cite{13}).

We find that the $k_T$-factorization approach reasonably well describe a full
set of experimental data taken at the Tevatron and LHC. Moreover, the 
shape and absolute normalization of measured cross sections are adequately reproduced
within the theoretical and experimental uncertainties.
At the Tevatron, there is some discrepancy between recent CDF data\cite{2} and our predictions at very high
diphoton invariant masses (see Fig.~7). Note, however, that the small-$x$
approximation used in our consideration is broken at high $M_{\gamma \gamma}$ 
and the more accurate 
treatment of, in particular, off-shell quark spin density matrix is needed.
This point, of course, is of importance but it is out of our present consideration.
From another side, we would like to point out that $k_T$-factorization approach provides a better 
description of all kinematic variables 
(in comparison with the NLO/NNLL pQCD calculations\cite{12,13}) 
at low $M_{\gamma \gamma}$ values, 
where expected effects connected with the small-$x$ physics 
becomes important. In the collinear QCD factorization,
observed large discrepancy between data and NLO/NNLL pQCD predictions\cite{12,13} at low $M_{\gamma \gamma}$
attributes to a fragmentation contributions and
higher-order QCD corrections to the gluon-gluon fusion subprocess\cite{1,2,3,4,5}.
However, the main part of these corrections are already effectively taken into 
account in our calculations. It is a radiative
QCD corrections described by the ladder-type diagrams 
which are naturally included into the $k_T$-factorization 
formalism at LO level\footnote{See review\cite{18} for more details.}.
At the LHC, our predictions agree well with the CMS data\cite{4} in a whole 
kinematical range and slightly overestimate
the ATLAS data\cite{5} at low values of $\Delta \phi_{\gamma \gamma}$ and
high values of $p_T^{\gamma \gamma}$. This point, probably, can indicate
the inconsistecy of the CMS and ATLAS data and, of course, needs in a further
theoretical and experimental investigations.
Note also that the NLO pQCD predictions\cite{12} underestimate the measured $\cos \theta^*$
distributions at both the Tevatron and LHC energies\cite{2,3,4,5}. This underestimation
is more significant for the central rapidity range ($|\eta| < 1.44$) at the LHC.
Our predictions agree well with the data\cite{2,3,4,5}. It 
could be important for experimental detection of Higgs boson signal
and further investigations of Higgs properties.

\begin{table}
\begin{center}
\begin{tabular}{|l|c|}
\hline
   & \\
  Source & $\sigma(p\bar p \to \gamma \gamma + X)$ [pb]\\
   & \\
\hline
   & \\
   $k_T$-factorization (KMR) & $29.9^{+7.4}_{-6.1}$ (scales)\\
   & \\
   NLO pQCD\cite{12} (\textsc{diphox}) & $27.3$ $^{+3.0}_{-2.3}$ (scales) $\pm 1.1$ (PDFs) \\
   & \\
   CMS data\cite{2} & $31.0 \pm 1.8$ (stat.) $^{+2.0}_{-2.1}$ (syst.) $\pm 1.2$ (lumi.)\\
   & \\
\hline
\end{tabular}
\end{center}
\caption{The total cross section of diphoton production in $pp$ collisions 
calculated in the kinematical region defined by $|\eta^{\gamma 1}| < 1.44$, 
$|\eta^{\gamma 2}| < 1.44$, $p_T^{\gamma 1} > 23$~GeV and 
$p_T^{\gamma 2} > 20$~GeV at $\sqrt s = 7$~TeV.}
\end{table}

As it was expected, the main important properties of the $k_T$-factorization approach
clearly manifest themselves in the calculated $p_T^{\gamma \gamma}$, $z$
and $\Delta \phi_{\gamma \gamma}$ distributions.
In the naive LO pQCD approximation, photons produced in $2\to 2$ subprocesses
are back-to-back in the transverse plane and
are balanced in $p_T$ due to momentum conservation. Therefore, these three ditsributions
must be simply a delta functions: $d\sigma/dp_T^{\gamma \gamma} \sim \delta(p_T^{\gamma \gamma})$,
$d\sigma/dz \sim \delta(z - 1)$ and 
$d\sigma/d\Delta \phi_{\gamma \gamma} \sim \delta(\Delta \phi_{\gamma \gamma} - \pi)$.
When higher-order QCD processes are considered, 
the presence of additional quarks and/or gluons in 
the final state allows these distributions to be more spread.
In the $k_T$-factorization approach, taking into account the
non-vanishing initial parton transverse momentum leads to the violation of back-to-back
kinematics even at LO approximation. 
So, despite the fact that one from contributed subprocesses
is the $2 \to 3$ subprocess
(that makes the difference between the $k_T$-factorization predictions and ones
from the collinear approximation of QCD not well pronounced), 
obtained perfect description of measured $\Delta \phi_{\gamma \gamma}$ 
distributions\cite{1,2,3,4,5} is notable. Specially we point out a reasonable 
description of the Tevatron and LHC data at low 
$\Delta \phi_{\gamma \gamma}$ values where the significant underestimation 
of the data by the NLO/NNLL pQCD calculations\cite{12,13} was observed (see also Figs.~2, 4 and 6).
The important role of such angular correlations
for understanding an interaction dynamics is well known\cite{18}.
In particular, as it was demonstrated in\cite{33}, these correlations strongly
depend on the unintegrated parton densities involved in the calculations
and can be used as a test to 
distinguish the different approaches of non-collinear parton 
evolution\footnote{We note, however, that at moment the 
KMR prescription is only one provide us with 
the unintegrated quark densities $f_q(x,{\mathbf k}_T^2,\mu^2)$ 
which can be used in a wide region of $x$ and ${\mathbf{k}_T^2}$ in
phenomenological studies. Thus, the problem of estimation of theoretical 
uncertainties connected with unintegrated parton densities is still open.}.
We would like to note also that the scale uncertainties of our 
calculations are quite small at $\Delta \phi_{\gamma \gamma} \sim 0$
and increases when $\Delta \phi_{\gamma \gamma} \to \pi$.
Similar to $\Delta \phi_{\gamma \gamma}$ distributions,
the calculated $p_T^{\gamma \gamma}$ distributions also directly
connected with the unintegrated quark and gluon distributions
due to transverse momentum conservation. Perfect overall agreement
of our calculations and the data on the $\Delta \phi_{\gamma \gamma}$ and 
$p_T^{\gamma \gamma}$ shows that the KMR prescription for evaluation
of unintegrated parton densities reproduces well 
the transverse momenta of initial quarks and gluons
in a probed kinematical region. 

\begin{table}
\begin{center}
\begin{tabular}{|l|c|}
\hline
   & \\
  Source & $\sigma(p\bar p \to \gamma \gamma + X)$ [pb]\\
   & \\
\hline
   & \\
   $k_T$-factorization (KMR) & $59.2^{+14.4}_{-12.1}$ (scales)\\
   & \\
   NLO pQCD\cite{12} (\textsc{diphox}) & $52.7$ $^{+5.8}_{-4.2}$ (scales) $\pm 2.0$ (PDFs) \\
   & \\
   CMS data\cite{2} & $62.4 \pm 3.6$ (stat.) $^{+5.3}_{-5.8}$ (syst.) $\pm 2.5$ (lumi.)\\
   & \\
\hline
\end{tabular}
\end{center}
\caption{The total cross section of diphoton production in $pp$ collisions 
calculated in the kinematical region defined by $|\eta^{\gamma 1}| < 2.5$, 
$|\eta^{\gamma 2}| < 2.5$ (except pseudo-rapidity
region $1.44 < \eta < 1.57$ for both photons), $p_T^{\gamma 1} > 23$~GeV and 
$p_T^{\gamma 2} > 20$~GeV at $\sqrt s = 7$~TeV.}
\end{table}

Special interest for the $k_T$-factorization phenomenology
is connected also with the events with low invariant mass and high
transverse momentum of produced photon pair. The CDF collaboration has studied
this kinematical region using a special condition $M_{\gamma \gamma} < p_T^{\gamma \gamma}$
in their experimental analysis\cite{2}. Within the mentioned above cuts on the
transverse momenta of the produced photons,
such measurements provide us with an 
additional test for unintegrated quark and gluon distributions $f_a(x,{\mathbf k}_T^2,\mu^2)$ 
at moderate and high ${\mathbf k}_T^2$ values. The
events with back-to-back kinematics are strongly suppressed in this case (see Fig.~10).
In contrast with the
NLO/NNLL pQCD calculations\cite{12,13}, the $k_T$-factorization predictions 
agree well with the CDF data.
An opposite condition (namely, $M_{\gamma \gamma} > p_T^{\gamma \gamma}$)
corresponds to the selection of events with kinematics similar to the decay 
of a heavy particle with low transverse momentum into a photon pair, 
such as, for example, $H \to \gamma \gamma$. 
Such events are better described\cite{2} by the NLO pQCD calculations\cite{12} than ones
with low invariant mass and high transverse momentum. Our 
predictions agree within the uncertainties with the data\cite{2}, but
overall agreement is a bit worse.

Concerning a relative contributions to the calculated diphoton production cross sections,
we find that quark-antiquark subprocess dominates at both the Tevatron and LHC
energies in the probed kinematical regions\footnote{We strongly disagree with the 
estimations\cite{34} of relative contributions from the gluon-gluon fusion and quark-antiquark annihilation
subprocesses performed for the Tevatron energy.}. The contribution from 
quark-gluon scattering is similar to gluon-gluon fusion one at
low diphoton invariant masses and overshoot gluon-gluon fusion contribution 
at high $M_{\gamma \gamma}$ values.
As it was expected\cite{4,5}, the role of both these subprocesses increases significantly at the LHC energies.
We note also that the contribution 
from $q\bar q \to \gamma \gamma g$ subprocess,
which is order of ${\cal O}(\alpha_s \alpha^2)$, is already taken into account 
in our calculations. To be precise, this contribution 
is effectively included to the quark-antiquark annihilation one 
due to the initial state gluon radiation (see also\cite{18}).
Of course, it is in a clear contrast with the collinear QCD factorization where these contributions 
should be calculated separately.

The predicted total cross sections for the Tevatron 
conditions are listed in Table~1 and for LHC conditions are listed in Tables 2 and 3,
respectively. Our results are slightly below the Tevatron data and NLO/NNLL pQCD
predictions\cite{12,13} but agree with them within the theoretical and experimental uncertainties.
However, at the LHC energy, our predictions overshoot the NLO pQCD ones\cite{12}
and are more close to the data. It could be connected with the small-$x$ effects
which becomes more important for the LHC conditions.

In conclusion, we would like to stress a number of important achievements demonstrated by
the $k_T$-factorization QCD approach. As a general feature, the $k_T$-factorization 
predictions are found to be
perfectly compatible with the available data, in particular, on the 
production of prompt photons, Drell-Yan lepton pairs, heavy quarks as well as various 
quarkonium states at modern colliders\cite{20,21,22,23,24,25,33,35,36}.
The $k_T$-factorization approach succeeds in describing the polarization
phenomena observed in $ep$ and $pp$ interactions (see, for example,\cite{35,36} 
and references therein). The latter is essentially related to the initial gluon 
off-shellness, which dominates the gluon polarization properties and has a considerable impact 
on the kinematics. 
So, we believe that the $k_T$-factorization formalism 
holds a possible key to understanding the production dynamics at high energies.

\section{Conclusions}

We have investigated the isolated prompt photon pair production in $p\bar p$ and $pp$ 
collisions at the Tevatron and LHC energies within the framework of the $k_T$-factorization QCD approach.
Our consideration is based on the quark-antiquark annihilation, quark-gluon scattering 
and gluon-gluon fusion subprocesses, where the non-zero transverse momenta of incoming
partons are taken into account.
The unintegrated parton densities in a proton are determined using the 
Kimber-Martin-Ryskin prescription. 
We obtained a reasonable well agreement between our predictions and recent data 
taken by the D$\emptyset$, CDF, CMS and ATLAS collaborations.
The achieved description of the data in several kinematical regions (in particular,
at low azimuthal angle distance between produced photons)
is better as compared to the one obtained in the framework of collinear QCD factorization (with NLO or 
even NNLL accuracy). 
Also, we describe reasonably well the Tevatron and LHC data on the $\cos \theta^*$ distributions
of the diphoton production.
Such quantities (underestimated by the NLO pQCD calculations) probe the 
angular momentum of the final state and different 
for QCD-mediated production as compared to the decay of a scalar Higgs boson.
It could be important for experimental detection of Higgs
and further theoretical and experimental investigations in this field.
We find that the scale uncertainties of our 
calculations are quite small at $\Delta \phi_{\gamma \gamma} \sim 0$, 
although they are exceed, of course, the uncertainties of 
NLO/NNLL pQCD calculations in general. Moreover, 
the problem of estimation of our theoretical 
uncertainties (connected in part with the non-collinear parton evolution scheme) 
is still open and further theoretical attempts (in order to investigate the unintegrated
quark distributions more detail) are necessary to reduce this uncertainty.
It is important for further studies of small-$x$ physics at hadron colliders, 
and, in particular, for searches of effects of new physics beyond the SM at the LHC.

\section{Acknowledgments} \indent 
The author would like to thank N.P.~Zotov,
S.P.~Baranov and H.~Jung for their encouraging interest 
and helpful discussions. Author is 
also grateful to DESY Directorate for the support in the 
framework of Moscow --- DESY project on Monte-Carlo
implementation for HERA --- LHC.
This research was supported in part by the 
FASI of Russian Federation (grant NS-3920.2012.2),
RFBR grants 11-02-01454-a and 12-02-31030,
grant of president of Russian Federation (MK-3977.2011.2) and the 
grant of Ministry of education and sciences of Russia (agreement 8412).

\newpage

\begin{figure}
\begin{center}
\epsfig{figure=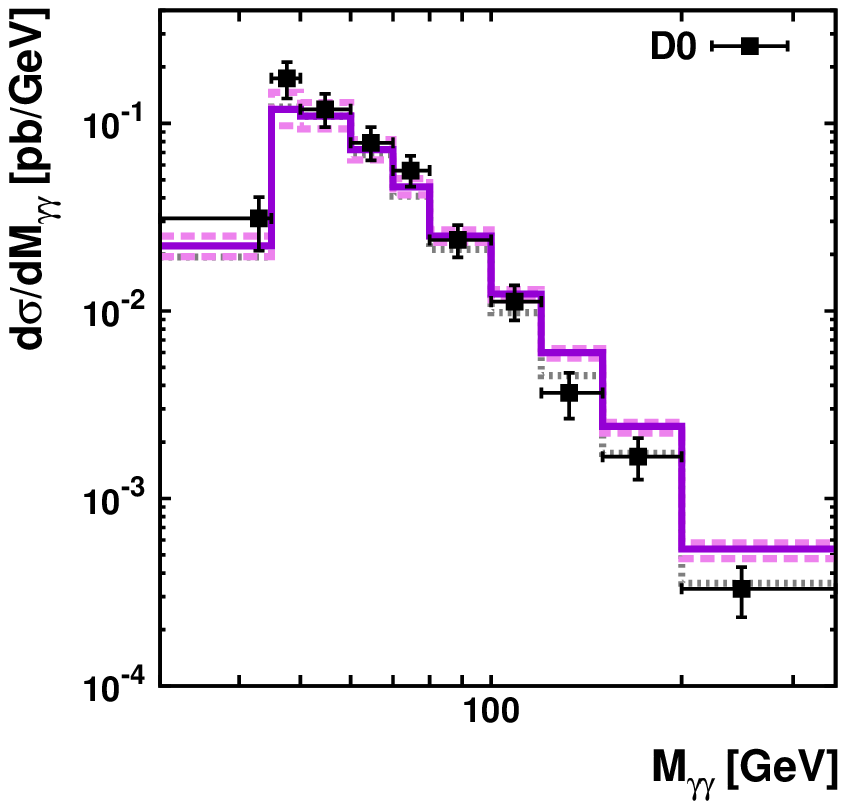, width = 8.1cm}
\epsfig{figure=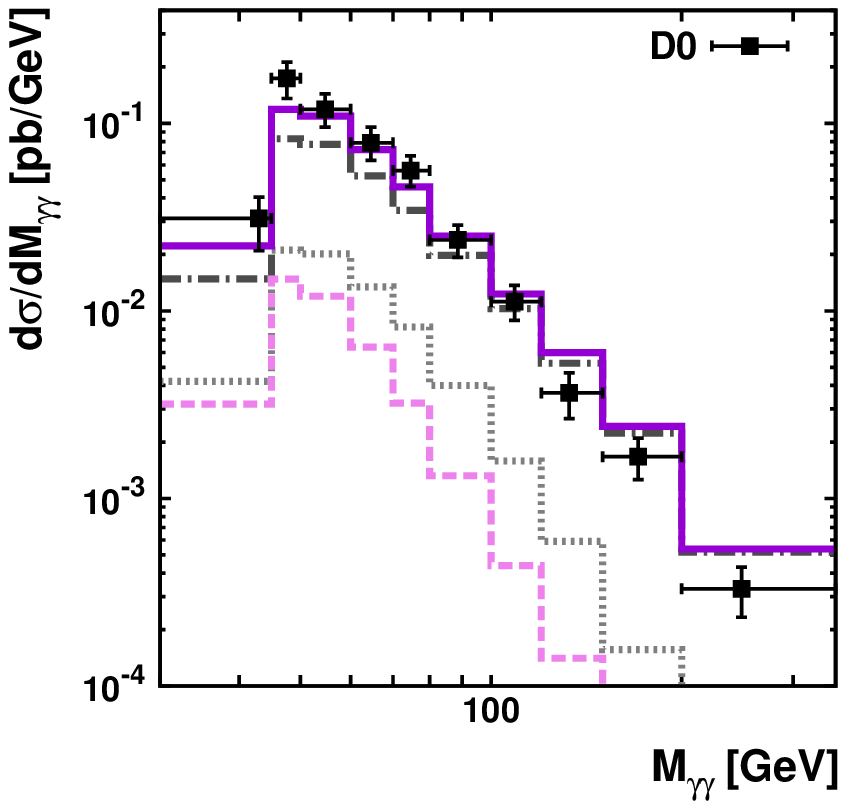, width = 8.1cm}
\epsfig{figure=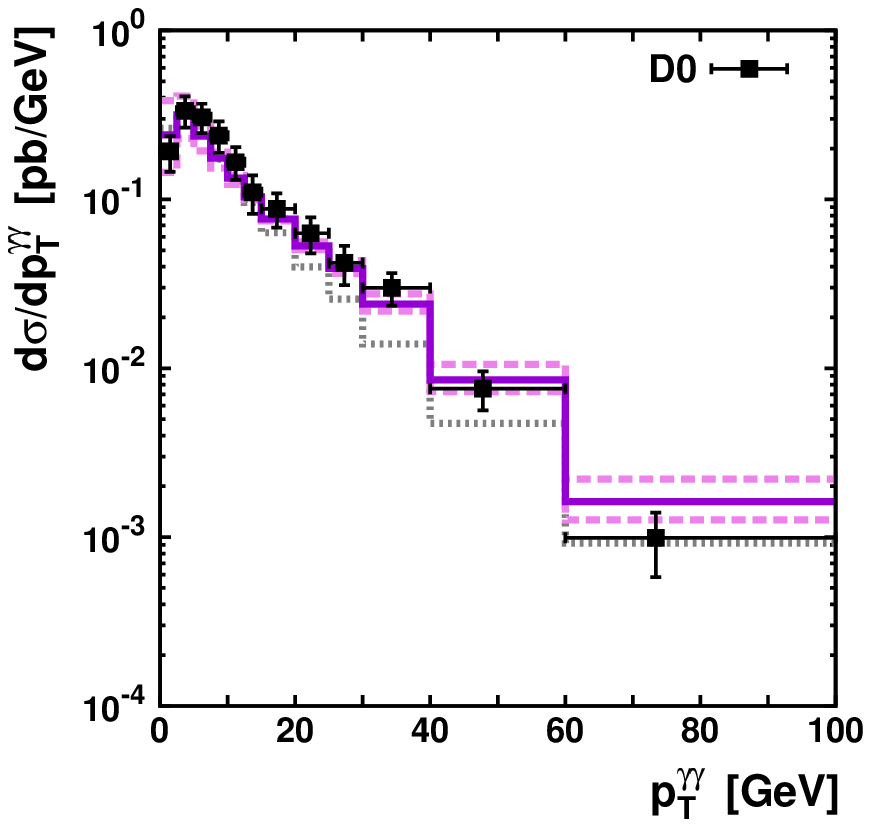, width = 8.1cm}
\epsfig{figure=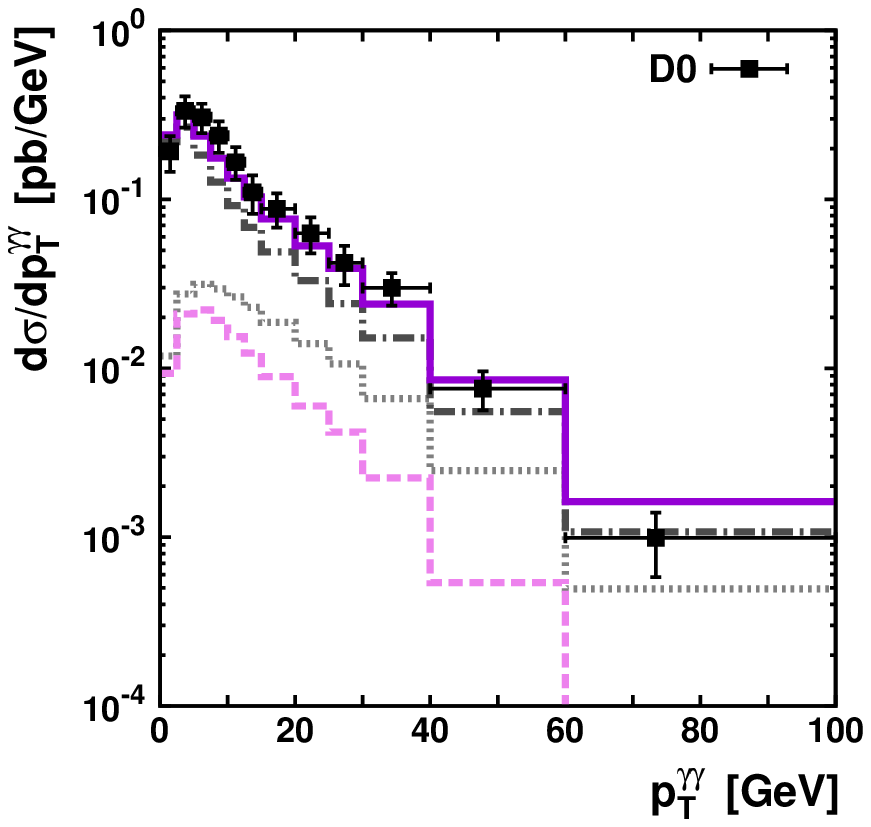, width = 8.1cm}
\caption{The differential cross section of  
prompt photon pair production in $p\bar p$ collisions
at the Tevatron as a function of diphoton invariant mass $M_{\gamma \gamma}$ and 
transverse momentum $p_T^{\gamma \gamma}$.
Left panel: the solid histograms correspond to the results obtained with the KMR parton densities
at the default scale, and the upper and lower dashed histograms correspond to standard scale variations, as it is
described in the text. The dotted histograms represent the 
NNLL pQCD predictions\cite{13} (as given by \textsc{resbos} program) listed in\cite{3}.
Right panel: the different contributions to the 
diphoton cross sections calculated at the default scale. 
The dashed, dash-dotted and dotted histograms correspond to the
gluon-gluon fusion, quark-antiquark annihilation and quark-gluon scattering subprocesses.
The solid histograms represent the sum of these contributions.
The experimental data are from D$\emptyset$\cite{3}.}
\label{fig1}
\end{center}
\end{figure}

\begin{figure}
\begin{center}
\epsfig{figure=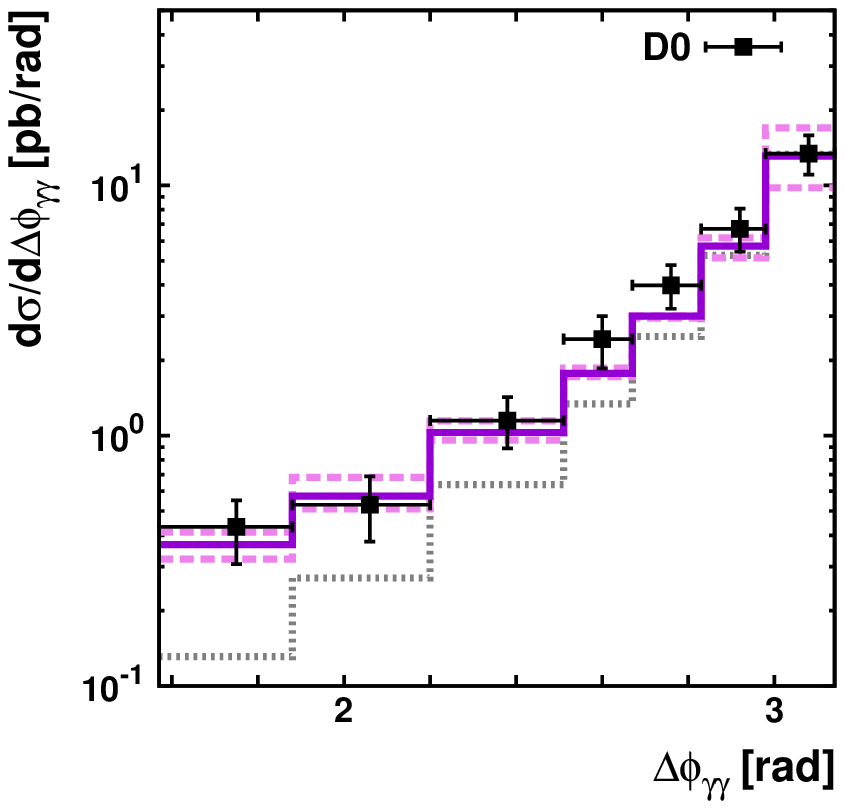, width = 8.1cm}
\epsfig{figure=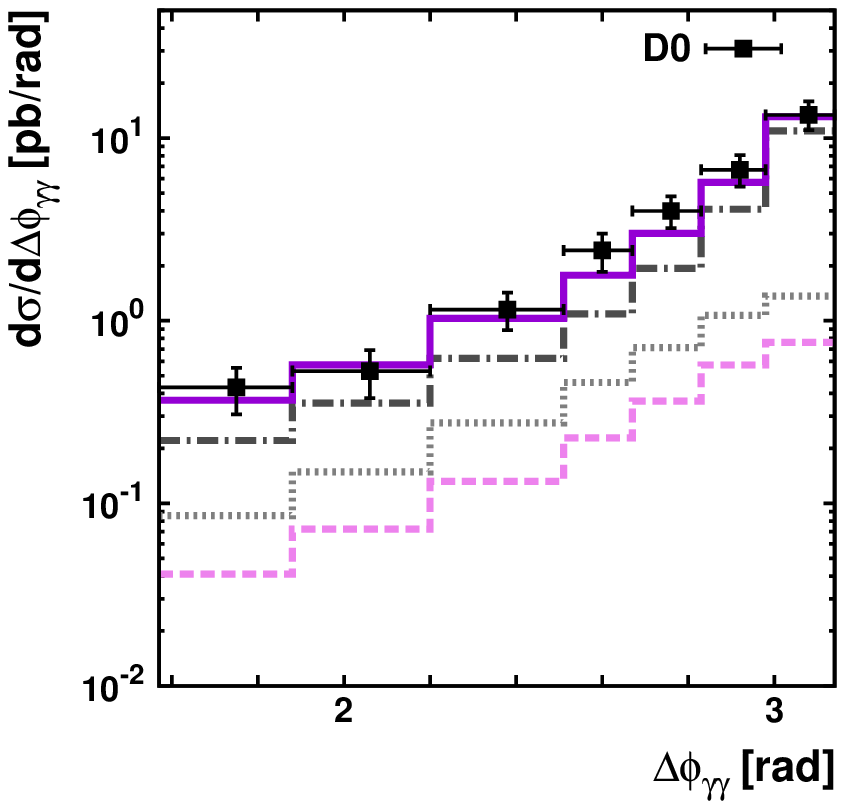, width = 8.1cm}
\epsfig{figure=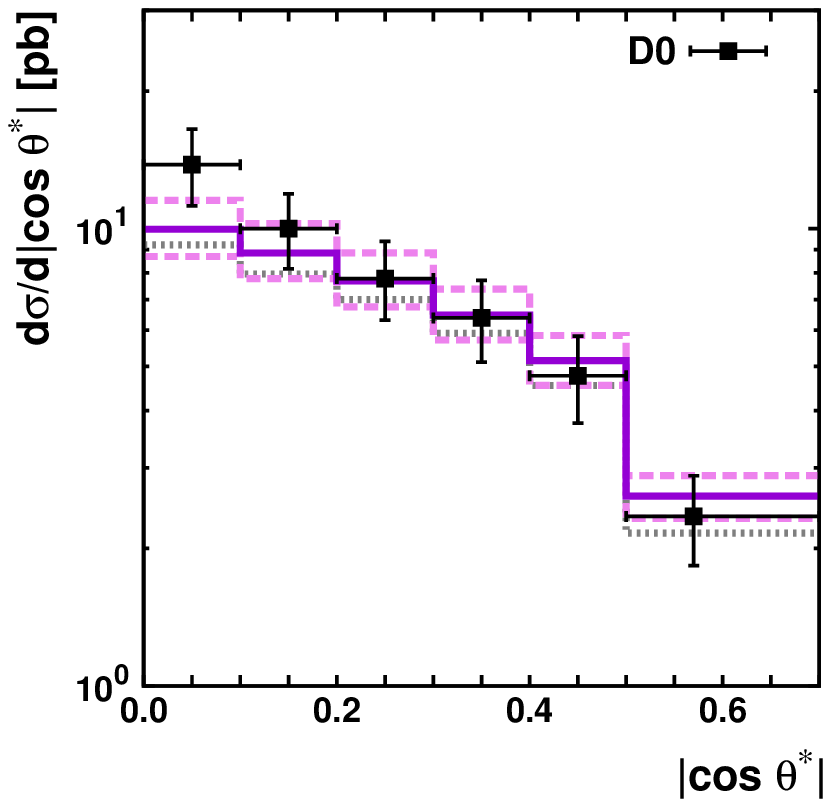, width = 8.1cm}
\epsfig{figure=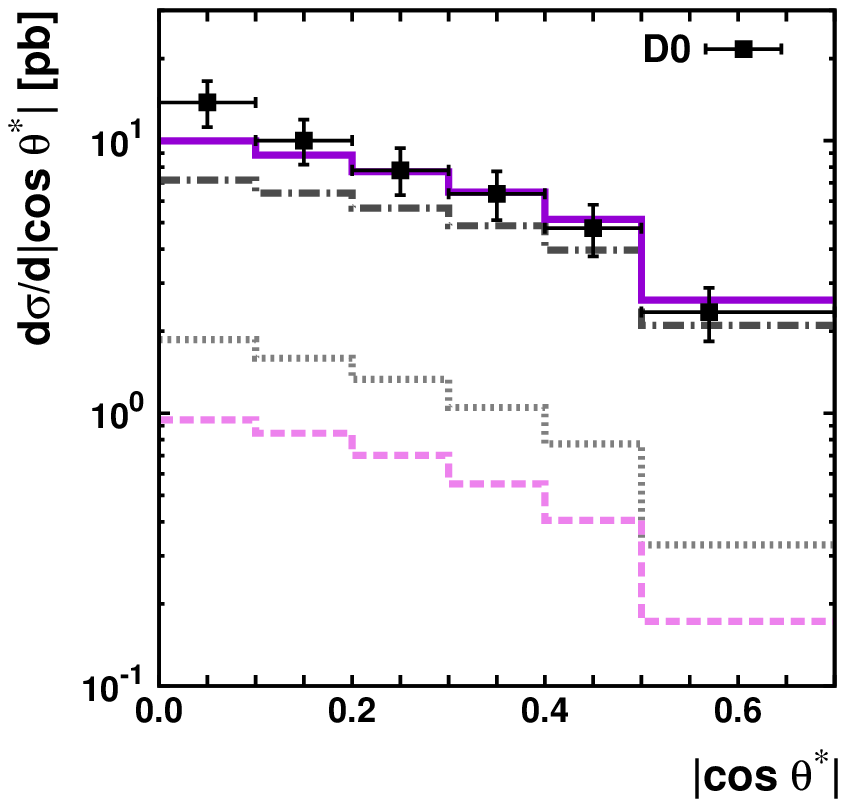, width = 8.1cm}
\caption{The differential cross section of prompt photon pair production in $p\bar p$ collisions
at the Tevatron as a function of $\Delta \phi_{\gamma \gamma}$ and $|\cos \theta^*|$.
Notation of all histograms is the same as in Fig.~2.
The experimental data are from D$\emptyset$\cite{3}.}
\label{fig2}
\end{center}
\end{figure}

\begin{figure}
\begin{center}
\epsfig{figure=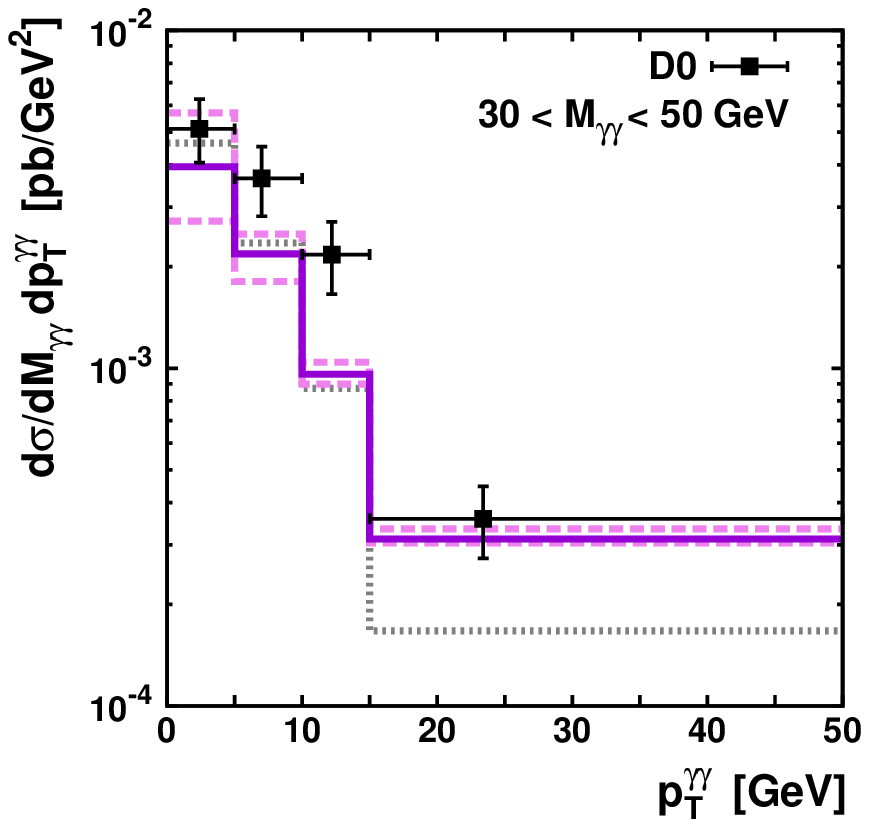, width = 8.1cm}
\epsfig{figure=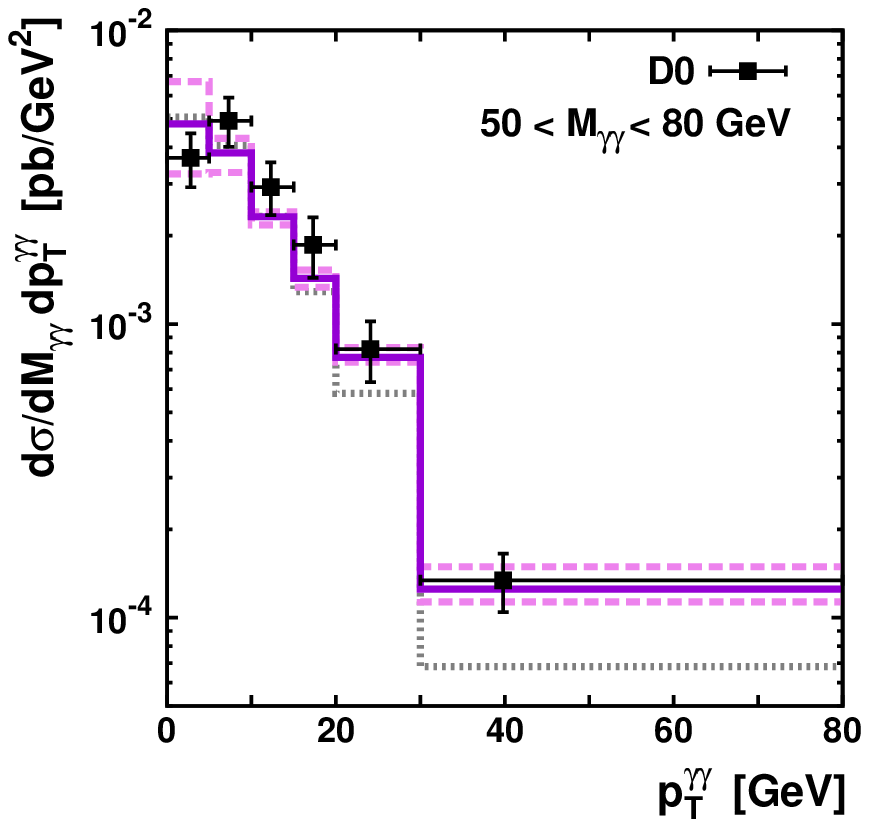, width = 8.1cm}
\epsfig{figure=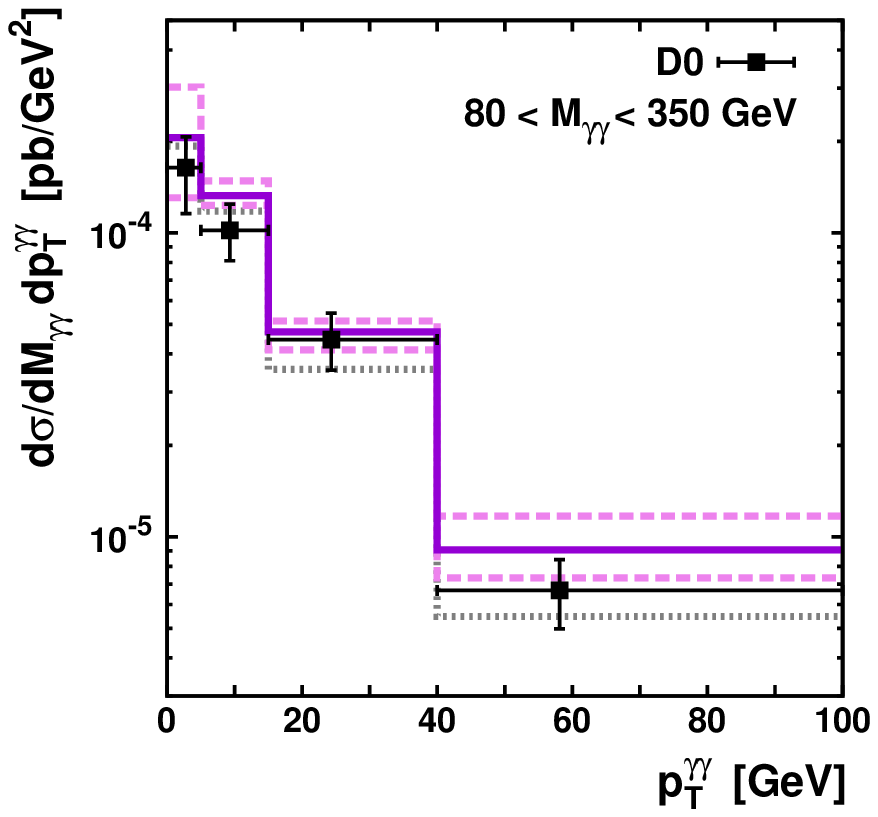, width = 8.1cm}
\caption{The double differential cross section 
$d\sigma/dM_{\gamma \gamma} dp_T^{\gamma \gamma}$
of prompt photon pair production in $p\bar p$ collisions
at the Tevatron calculated for three subdivisions of
$M_{\gamma \gamma}$ range.
The solid histograms correspond to the results obtained with the KMR parton densities
at the default scale, and the upper and lower dashed histograms correspond to standard scale 
variations, as it is described in the text. The dotted histograms represent the 
NNLL pQCD predictions\cite{13} (as given by \textsc{resbos} program) listed in\cite{3}.
The experimental data are from D$\emptyset$\cite{3}.}
\label{fig3}
\end{center}
\end{figure}

\begin{figure}
\begin{center}
\epsfig{figure=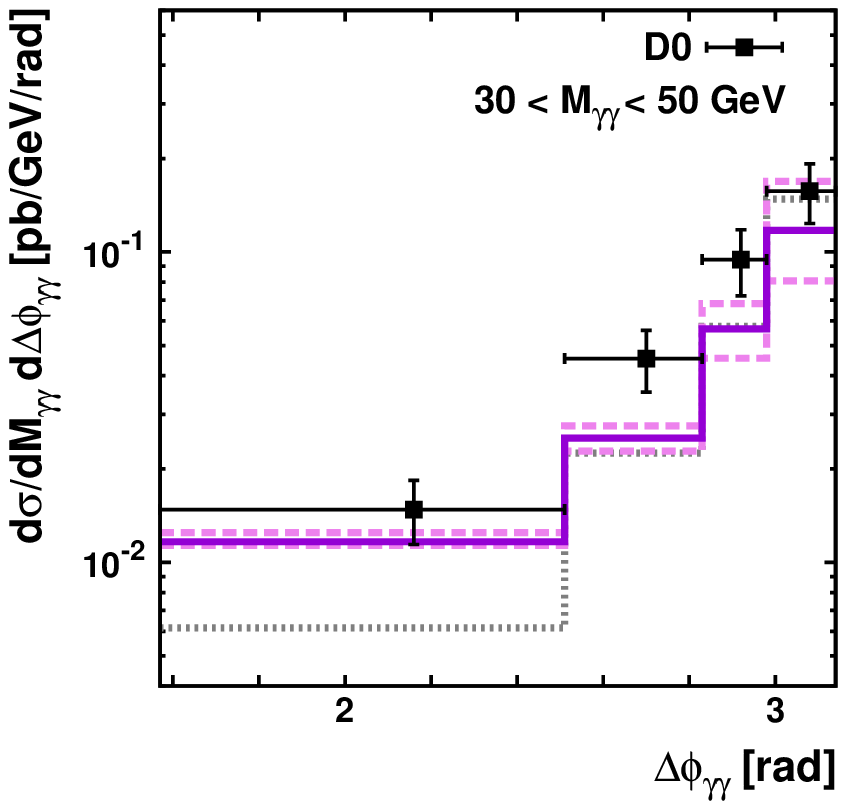, width = 8.1cm}
\epsfig{figure=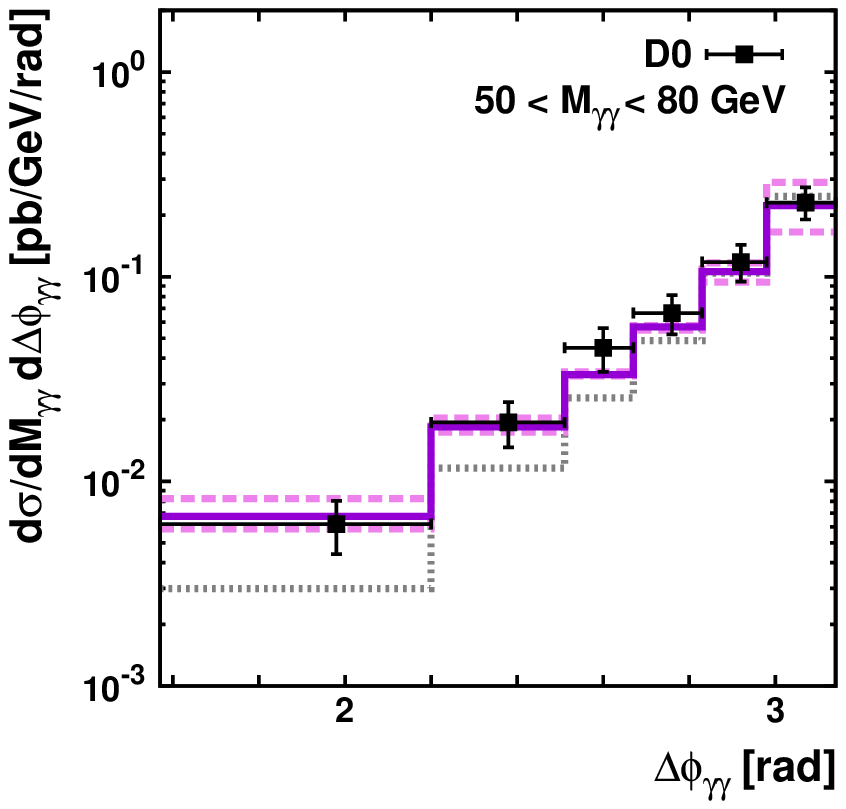, width = 8.1cm}
\epsfig{figure=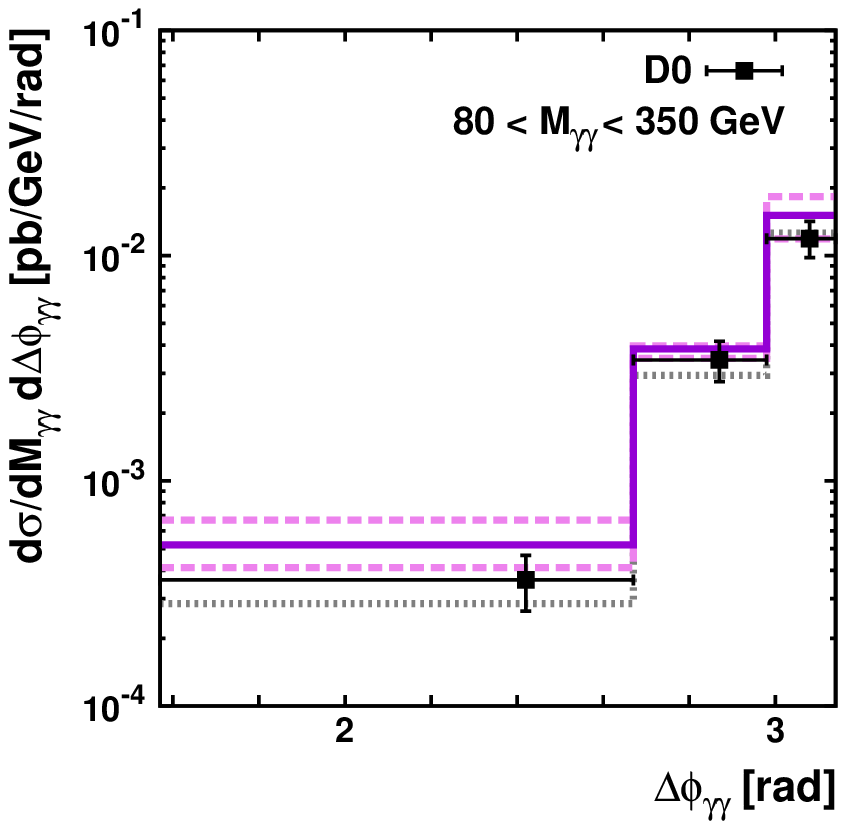, width = 8.1cm}
\caption{The double differential cross section 
$d\sigma/dM_{\gamma \gamma} d\Delta \phi_{\gamma \gamma}$
of prompt photon pair production in $p\bar p$ collisions
at the Tevatron calculated for three subdivisions of
$M_{\gamma \gamma}$ range.
Notation of all histograms is the same as in Fig.~3.
The experimental data are from D$\emptyset$\cite{3}.}
\label{fig4}
\end{center}
\end{figure}

\begin{figure}
\begin{center}
\epsfig{figure=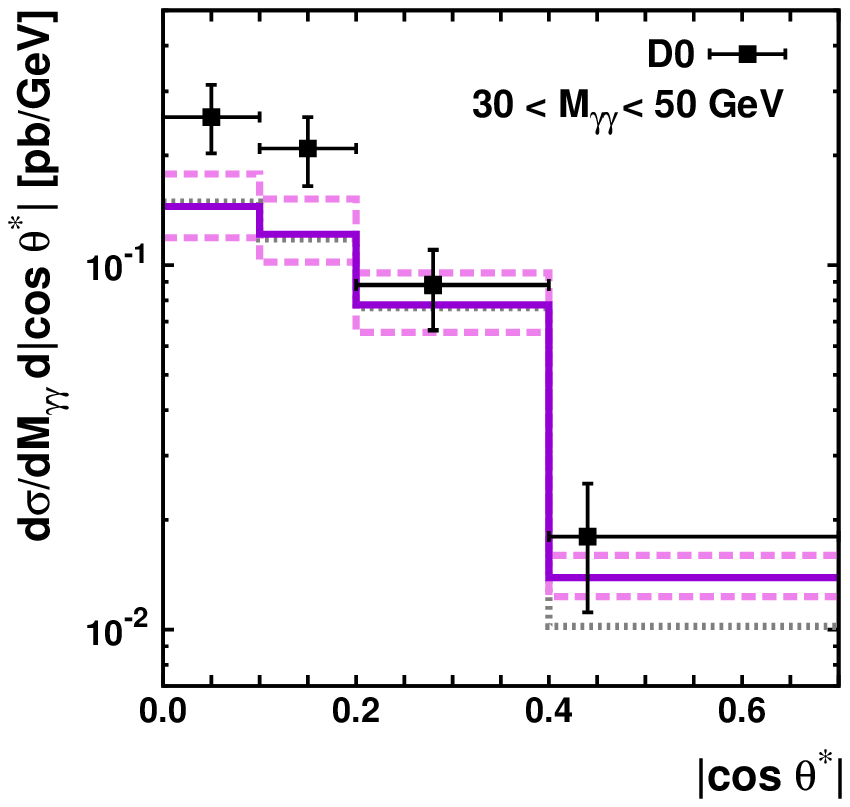, width = 8.1cm}
\epsfig{figure=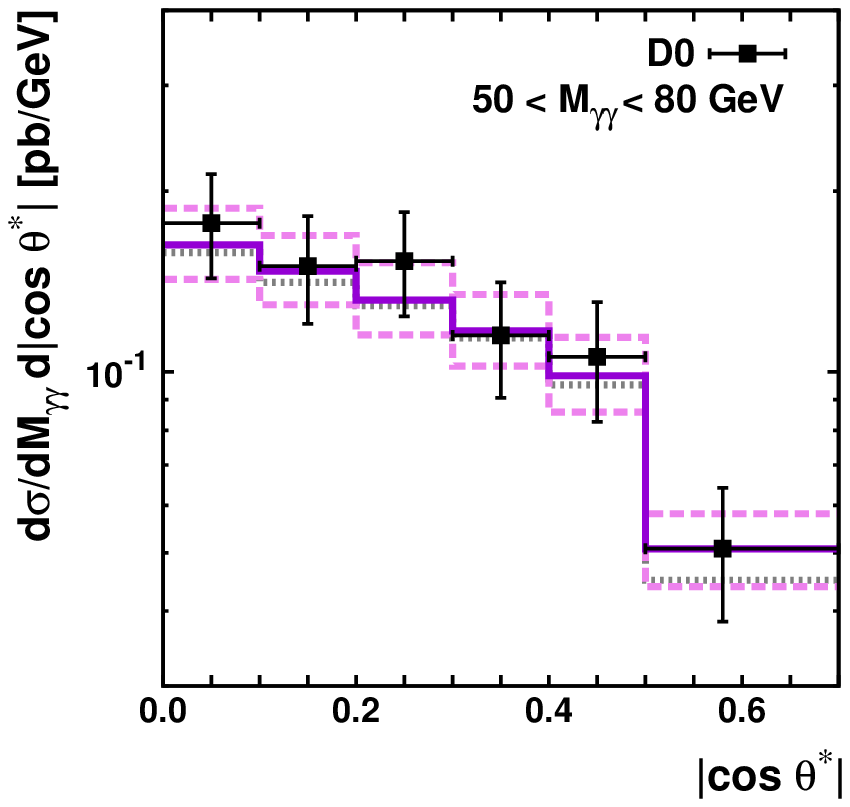, width = 8.1cm}
\epsfig{figure=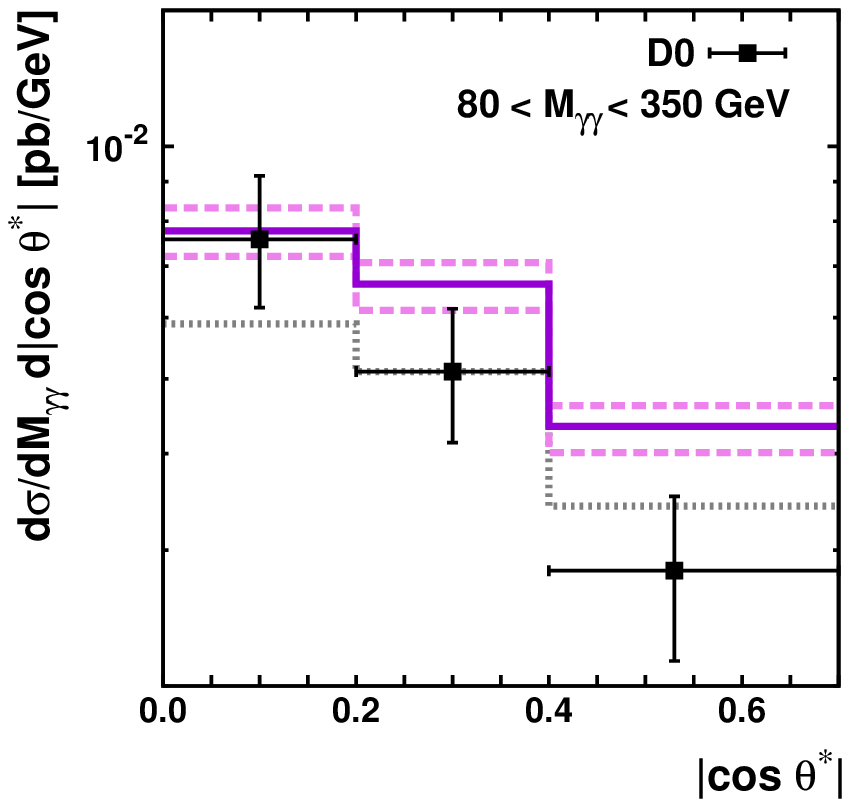, width = 8.1cm}
\caption{The double differential cross section 
$d\sigma/dM_{\gamma \gamma} d|\cos \theta^*|$
of prompt photon pair production in $p\bar p$ collisions
at the Tevatron calculated for three subdivisions of
$M_{\gamma \gamma}$ range.
Notation of all histograms is the same as in Fig.~3.
The experimental data are from D$\emptyset$\cite{3}.}
\label{fig5}
\end{center}
\end{figure}

\begin{figure}
\begin{center}
\epsfig{figure=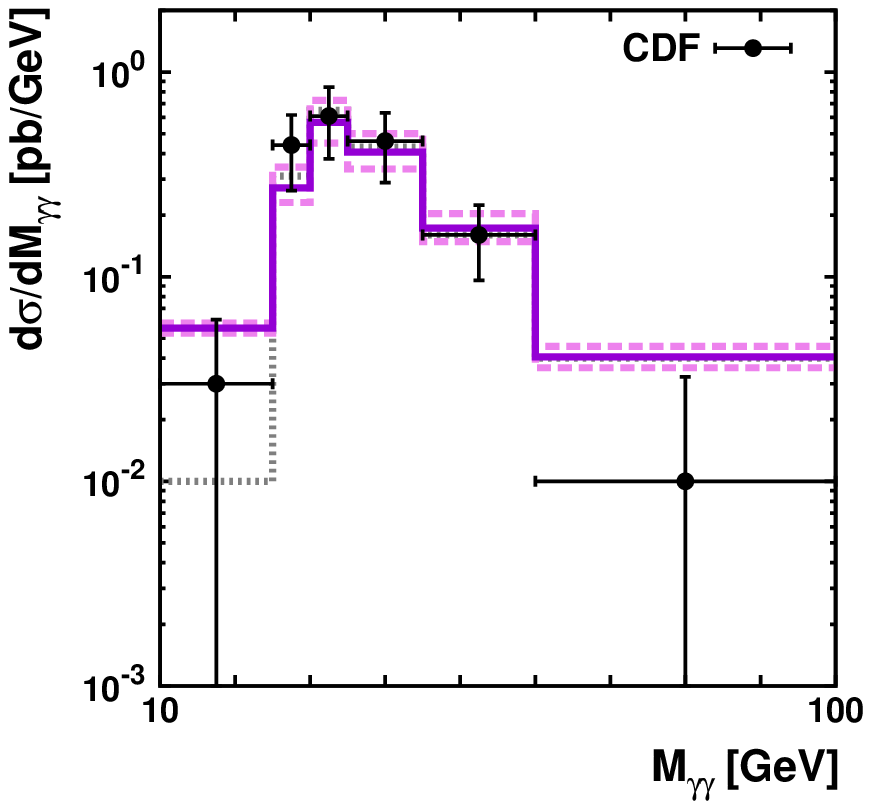, width = 8.1cm}
\epsfig{figure=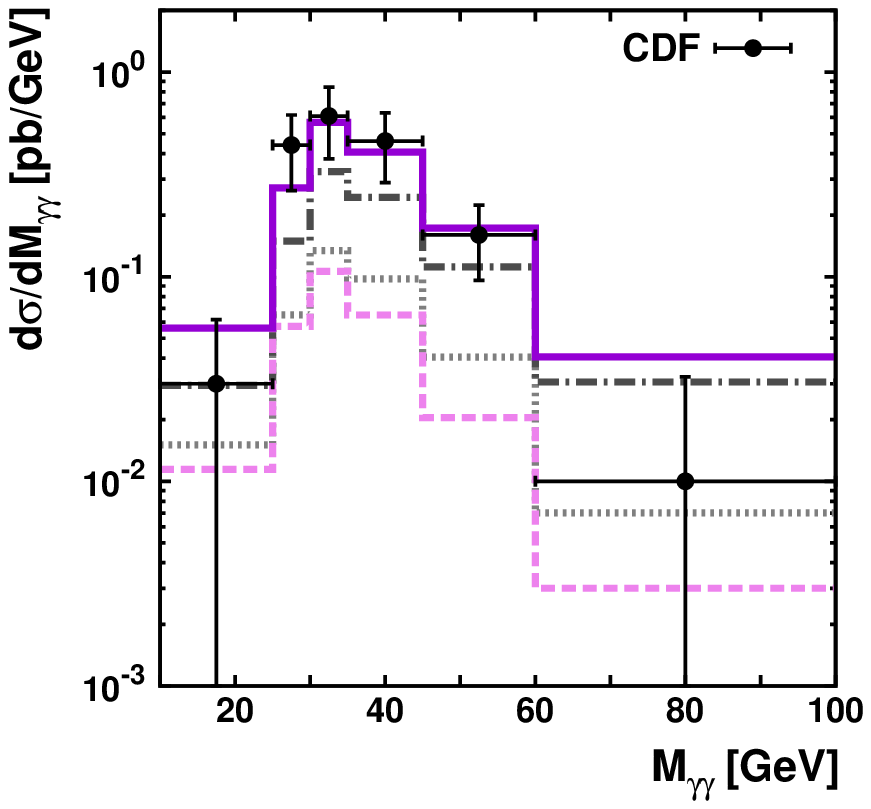, width = 8.1cm}
\epsfig{figure=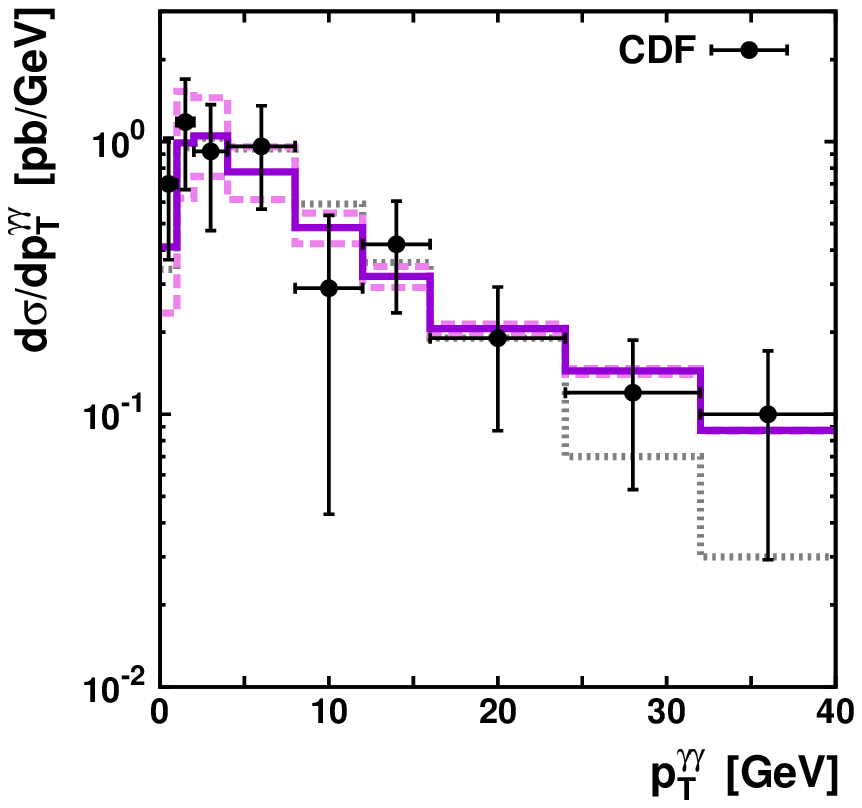, width = 8.1cm}
\epsfig{figure=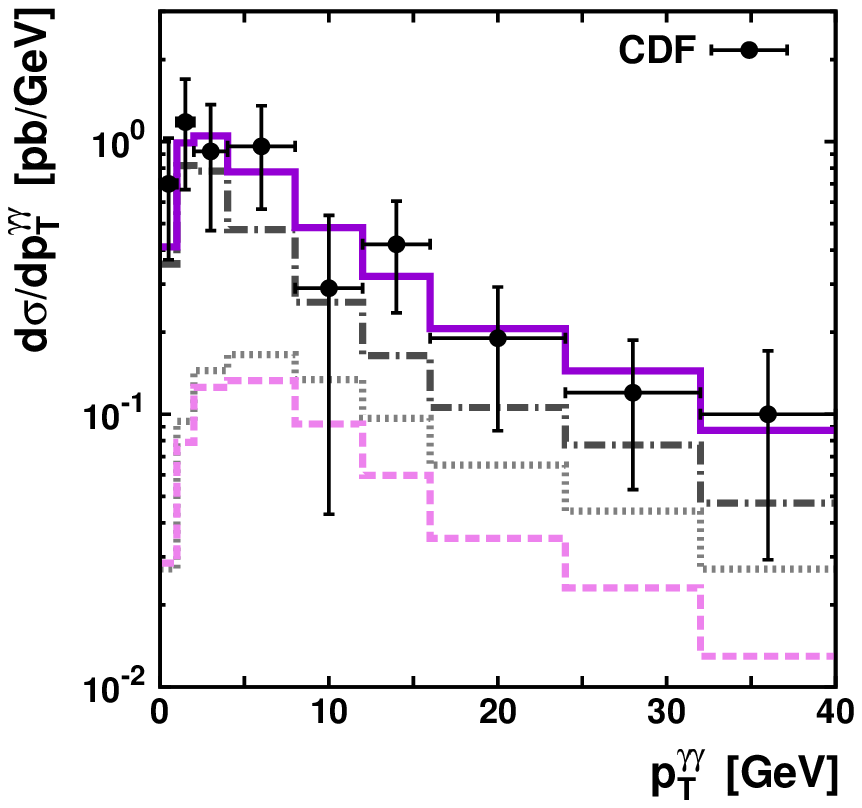, width = 8.1cm}
\epsfig{figure=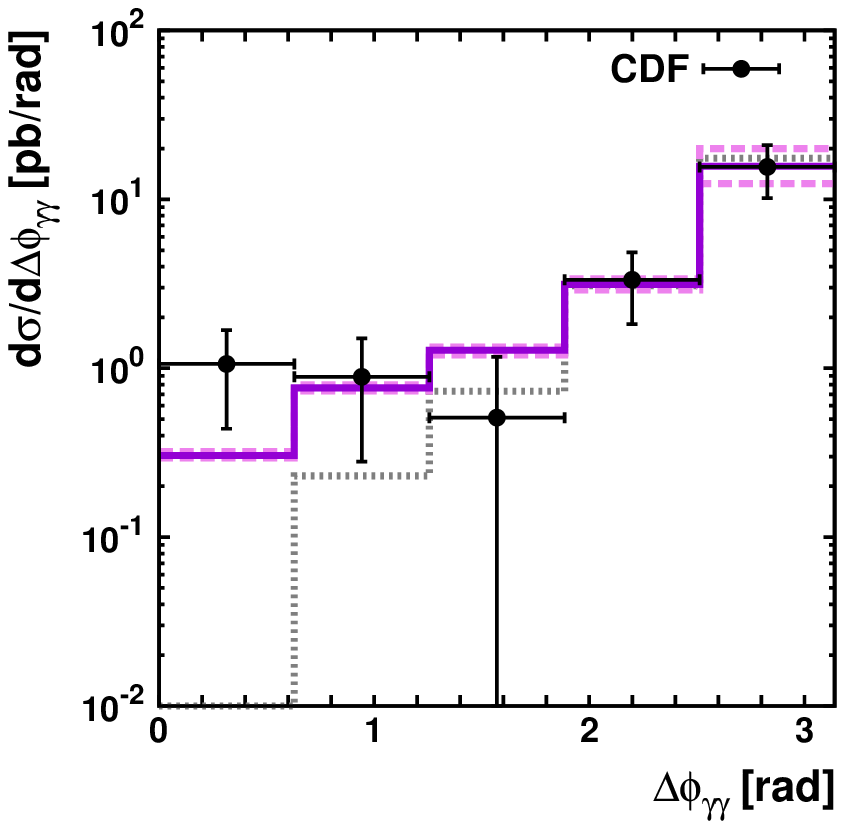, width = 8.1cm}
\epsfig{figure=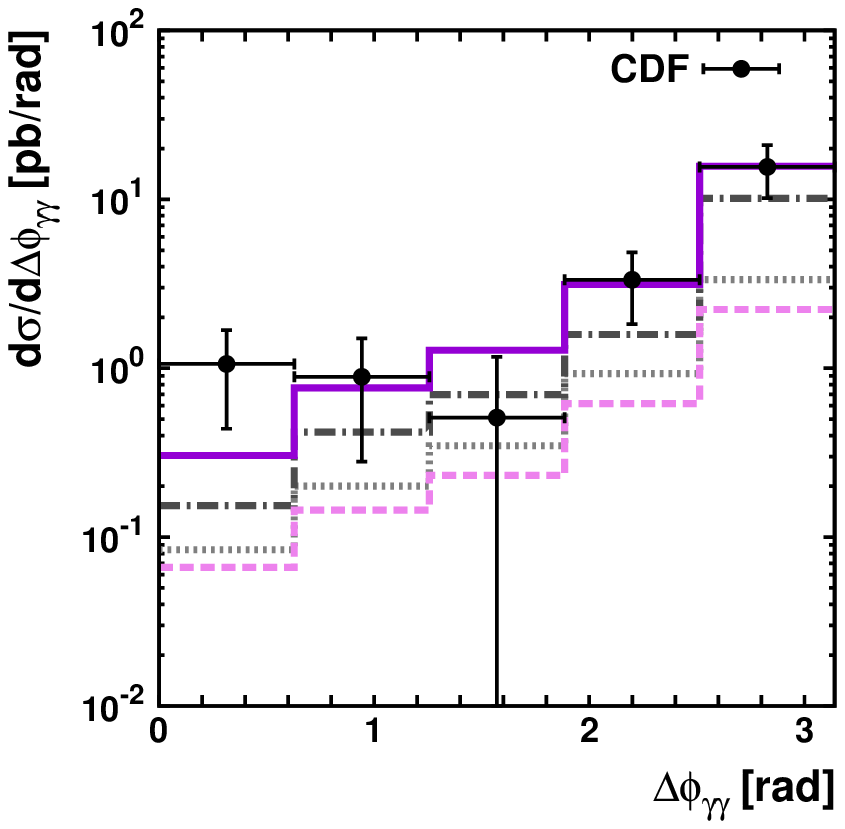, width = 8.1cm}
\caption{The differential cross section of  
prompt photon pair production in $p\bar p$ collisions
at the Tevatron.
Notation of all histograms is the same as in Fig.~2.
The experimental data are from CDF\cite{1}.}
\label{fig6}
\end{center}
\end{figure}

\begin{figure}
\begin{center}
\epsfig{figure=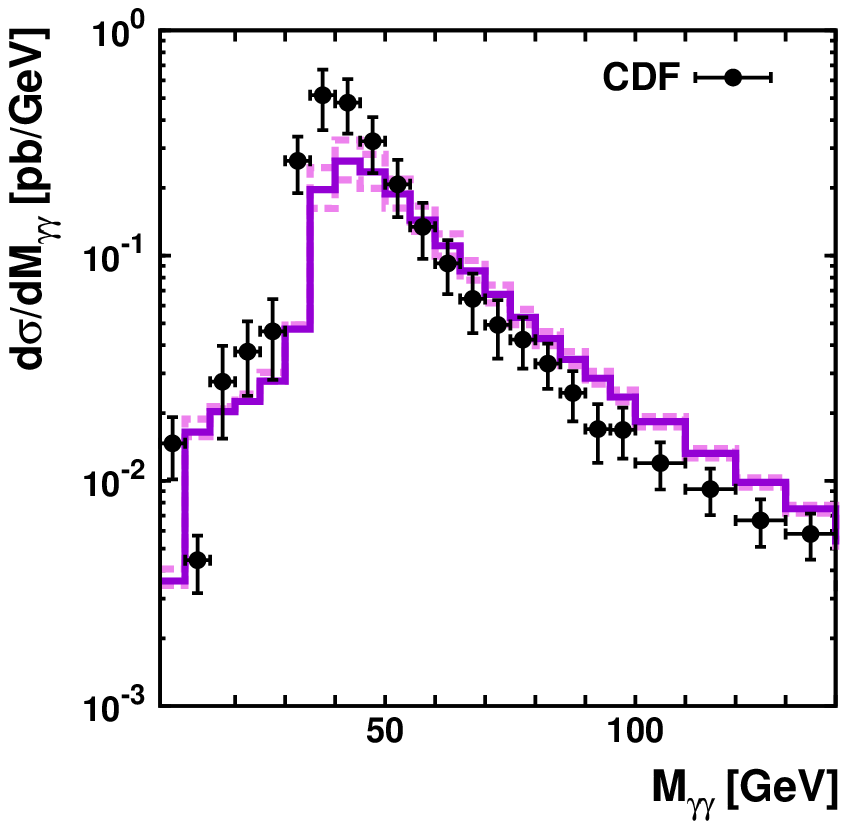, width = 8.1cm}
\epsfig{figure=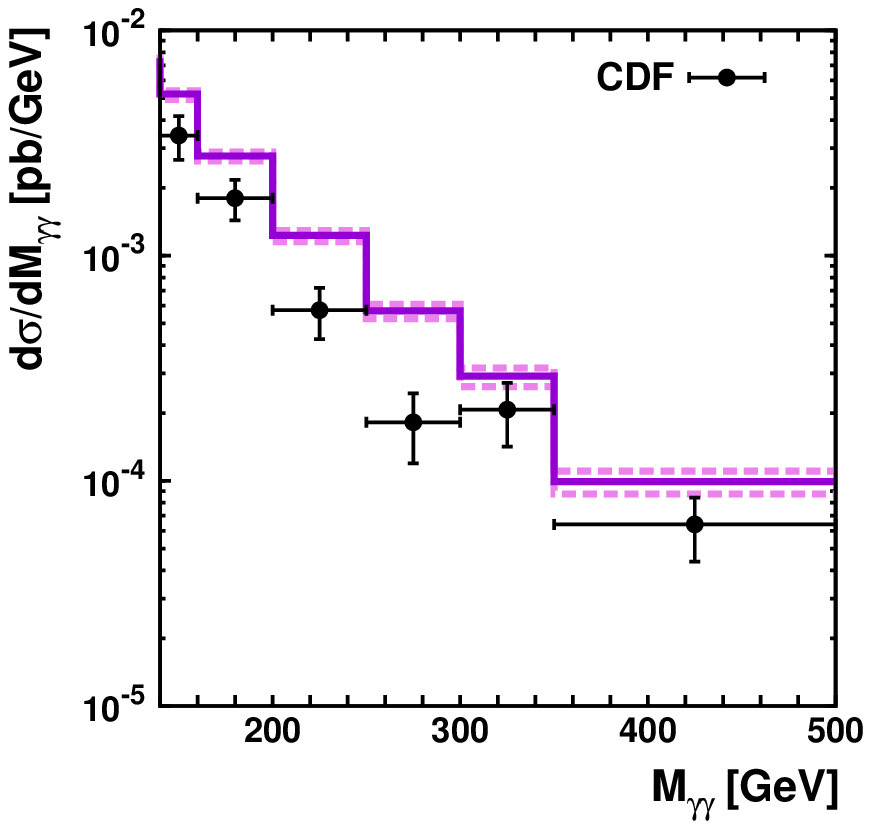, width = 8.1cm}
\epsfig{figure=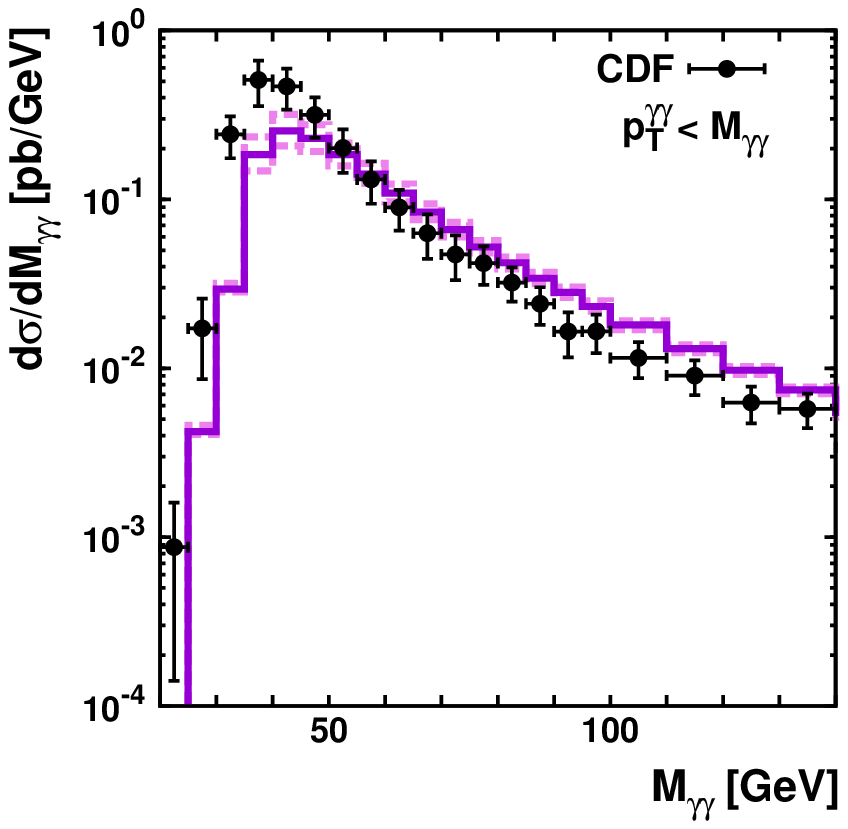, width = 8.1cm}
\epsfig{figure=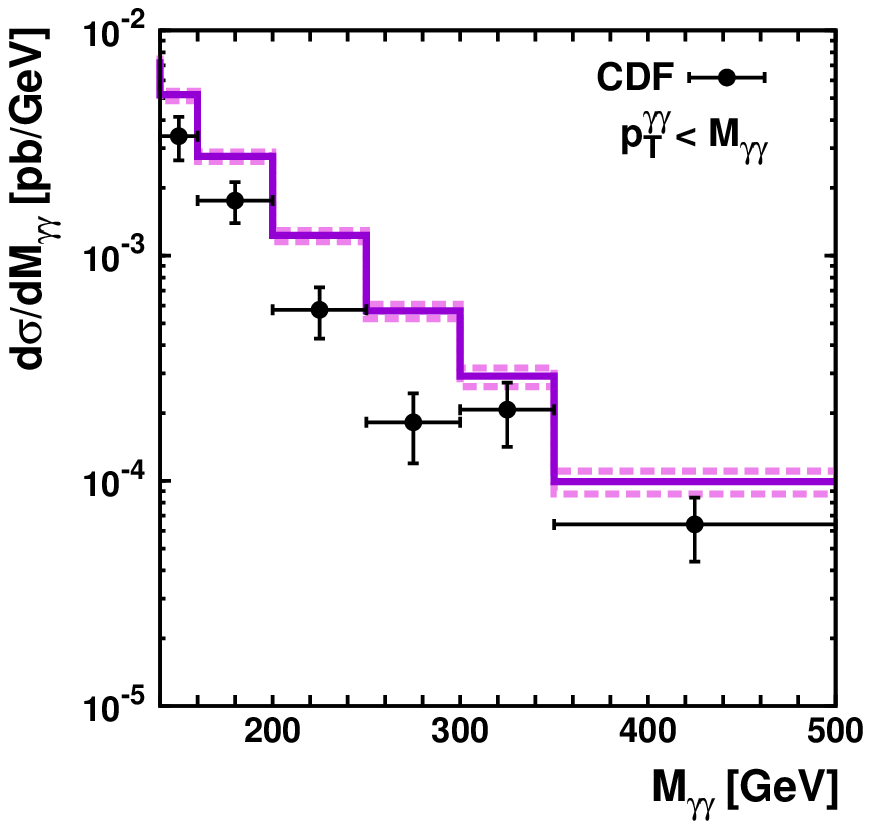, width = 8.1cm}
\epsfig{figure=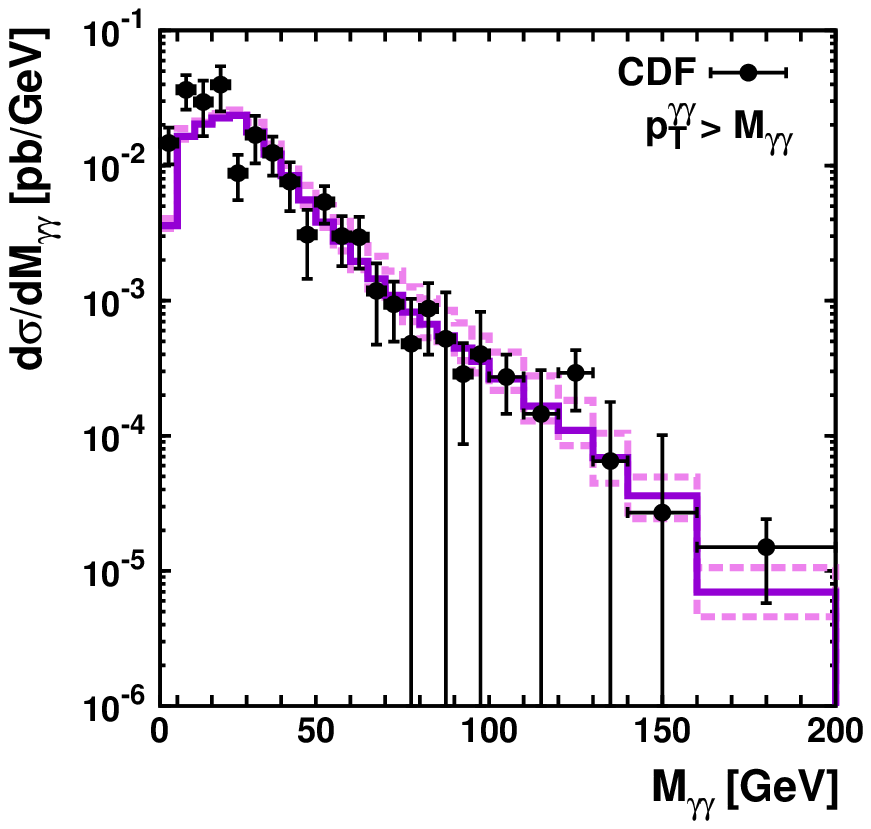, width = 8.1cm}
\caption{The differential cross section of  
prompt photon pair production in $p\bar p$ collisions
at the Tevatron as a function of diphoton invariant mass $M_{\gamma \gamma}$.
Two kinematic cases are shown separately:
differential cross sections for $M_{\gamma \gamma} > p_T^{\gamma \gamma}$ and
$M_{\gamma \gamma} < p_T^{\gamma \gamma}$.
The solid histograms correspond to the results obtained with the KMR parton densities
at the default scale, and the upper and lower dashed histograms correspond to standard scale 
variations, as it is described in the text.
The experimental data are from CDF\cite{2}.}
\label{fig7}
\end{center}
\end{figure}

\begin{figure}
\begin{center}
\epsfig{figure=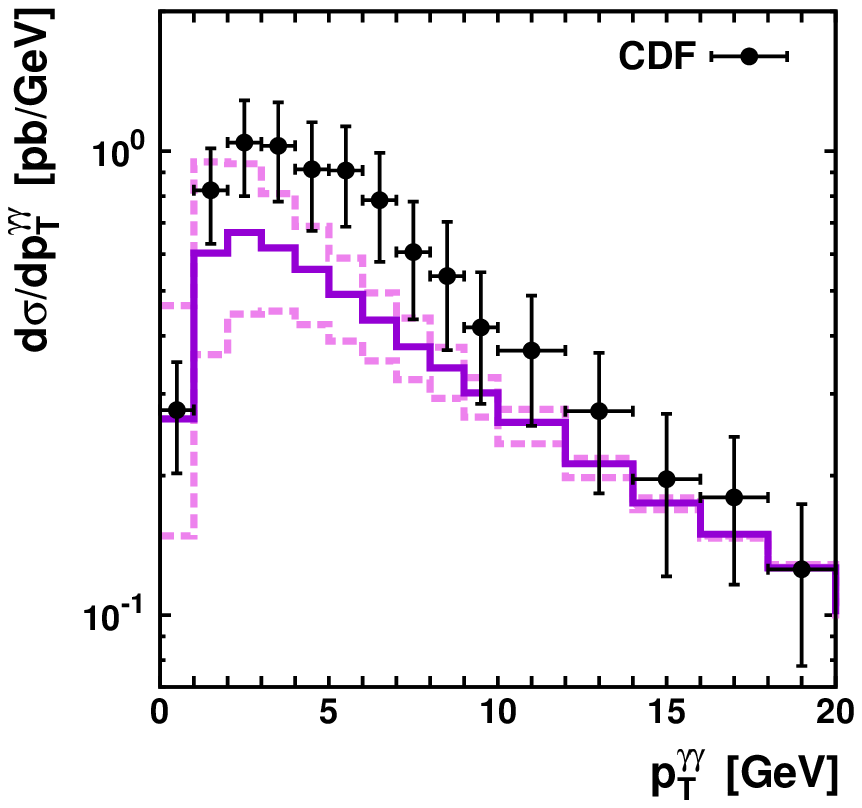, width = 8.1cm}
\epsfig{figure=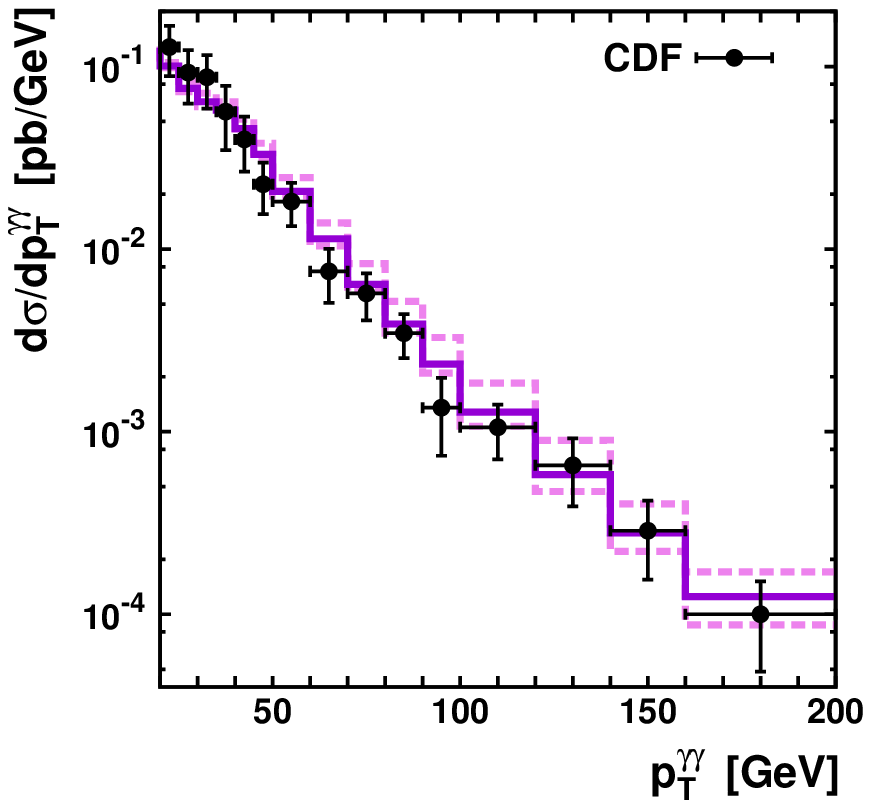, width = 8.1cm}
\epsfig{figure=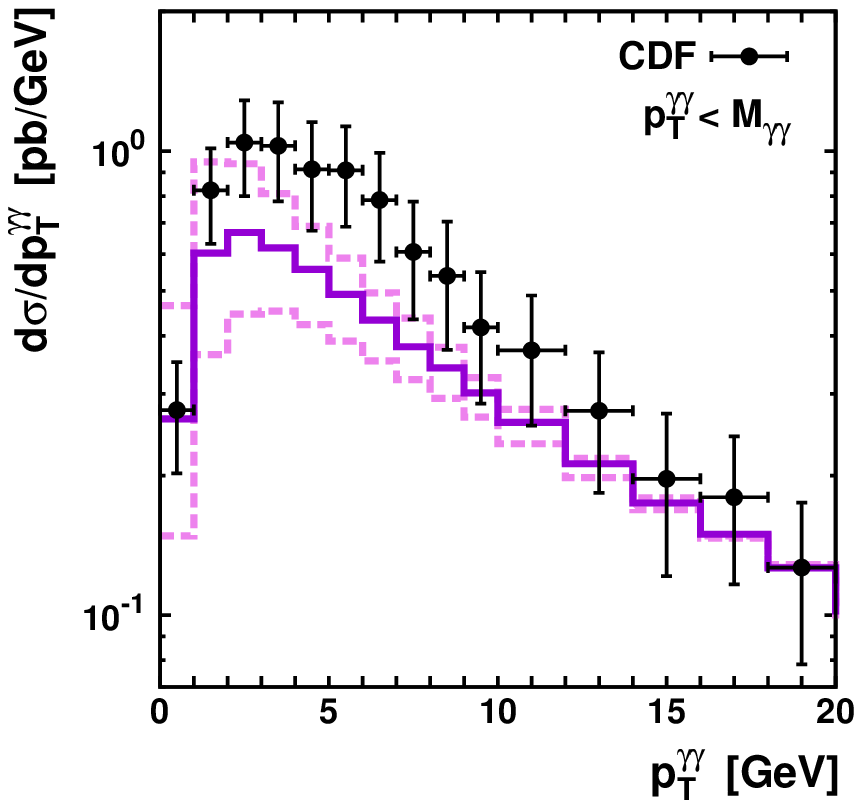, width = 8.1cm}
\epsfig{figure=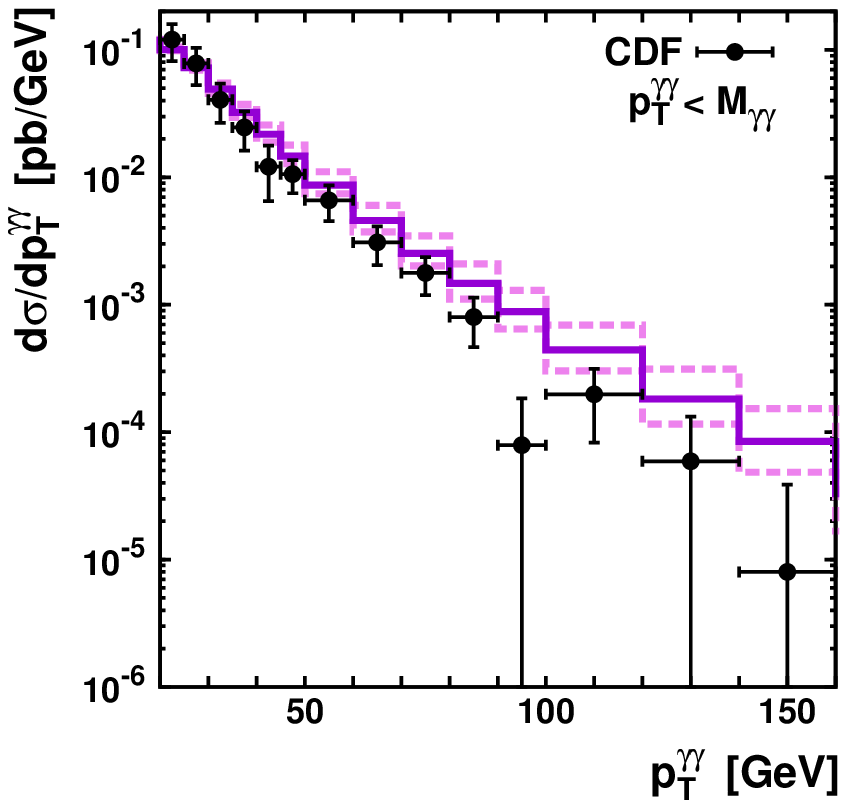, width = 8.1cm}
\epsfig{figure=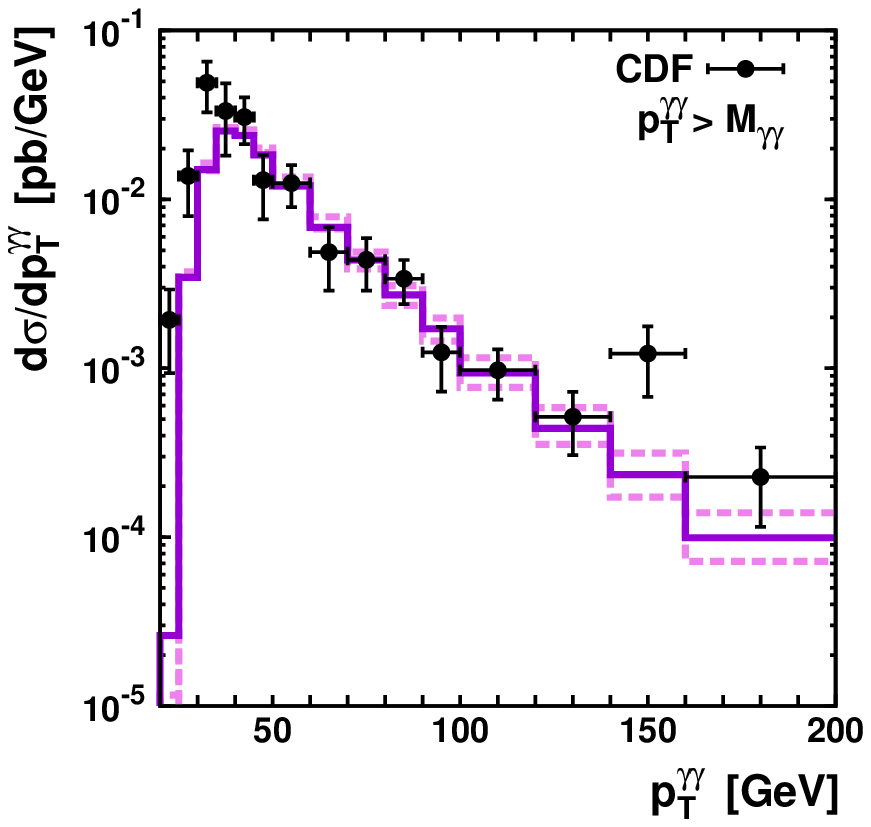, width = 8.1cm}
\caption{The differential cross section of  
prompt photon pair production in $p\bar p$ collisions
at the Tevatron as a function of diphoton transverse momentum $p_T^{\gamma \gamma}$.
Two kinematic cases are shown separately:
differential cross sections for $M_{\gamma \gamma} > p_T^{\gamma \gamma}$ and
$M_{\gamma \gamma} < p_T^{\gamma \gamma}$.
Notation of all histograms is the same as in Fig.~7.
The experimental data are from CDF\cite{2}.}
\label{fig8}
\end{center}
\end{figure}

\begin{figure}
\begin{center}
\epsfig{figure=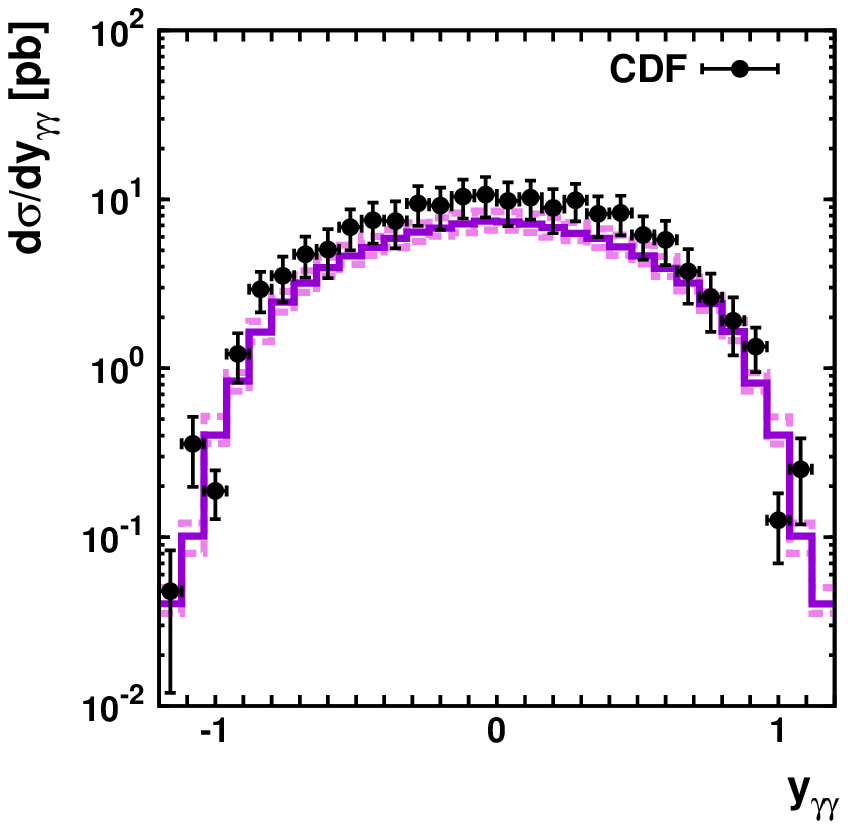, width = 8.1cm}
\epsfig{figure=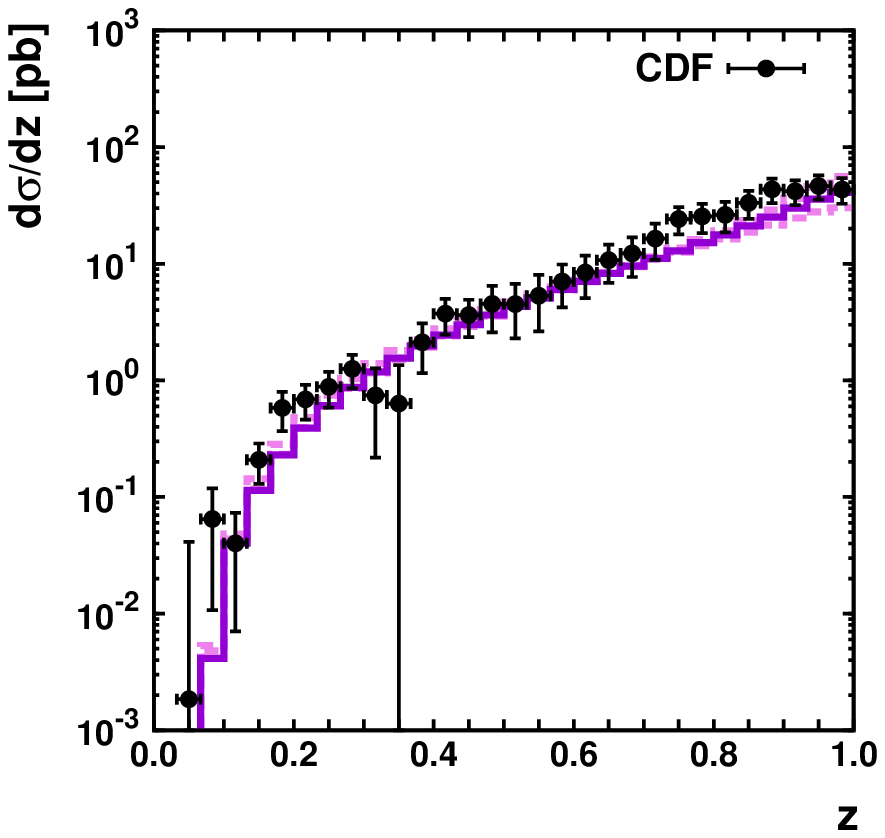, width = 8.1cm}
\epsfig{figure=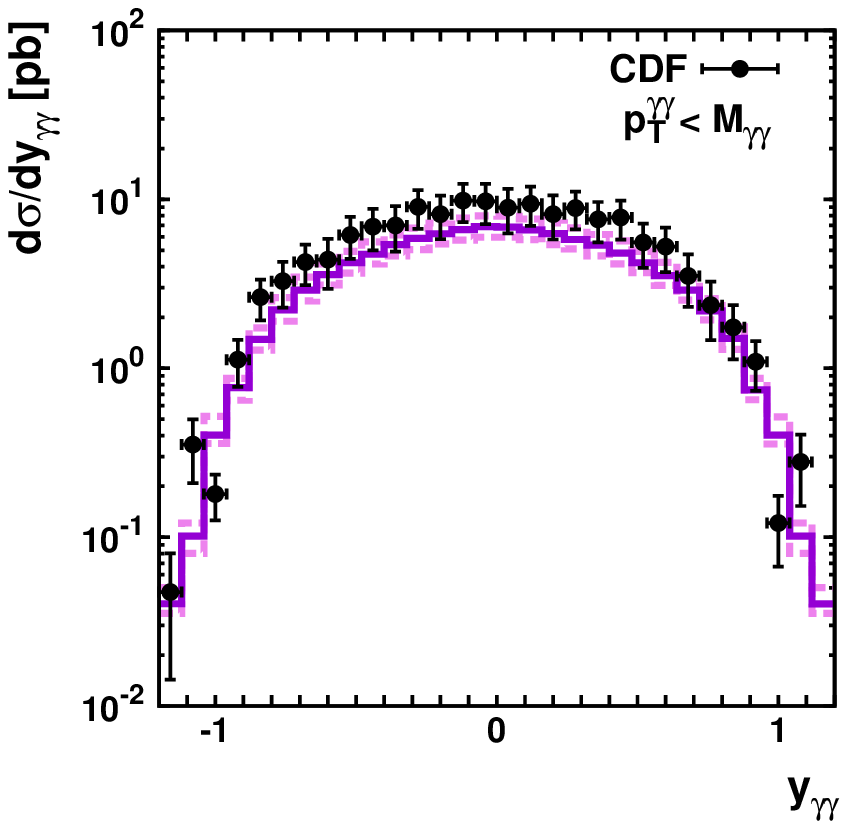, width = 8.1cm}
\epsfig{figure=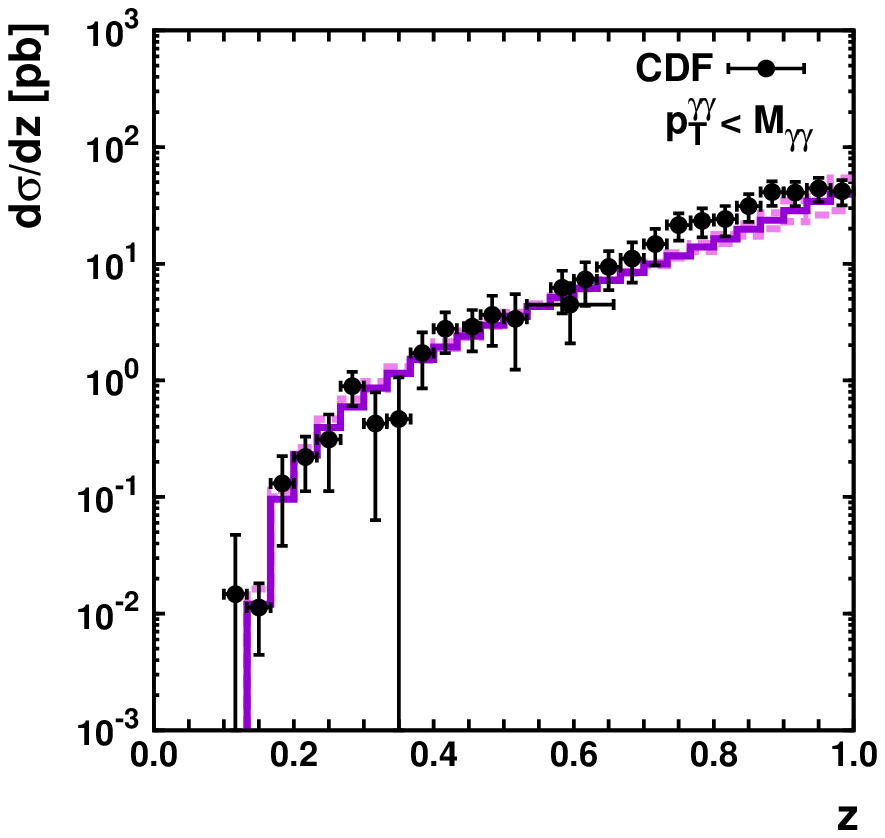, width = 8.1cm}
\epsfig{figure=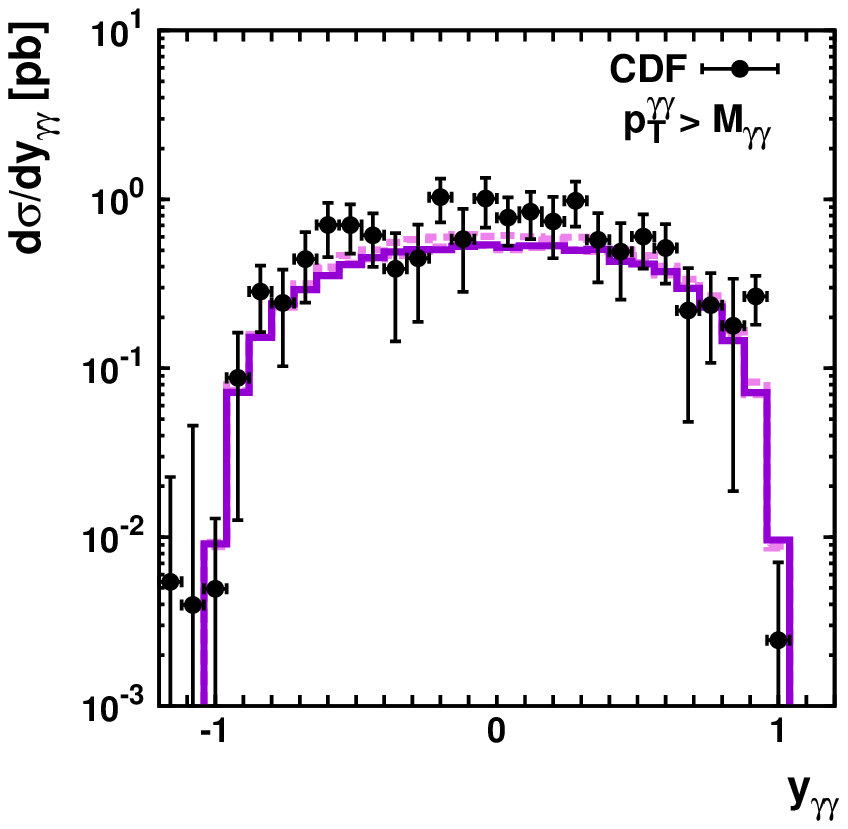, width = 8.1cm}
\epsfig{figure=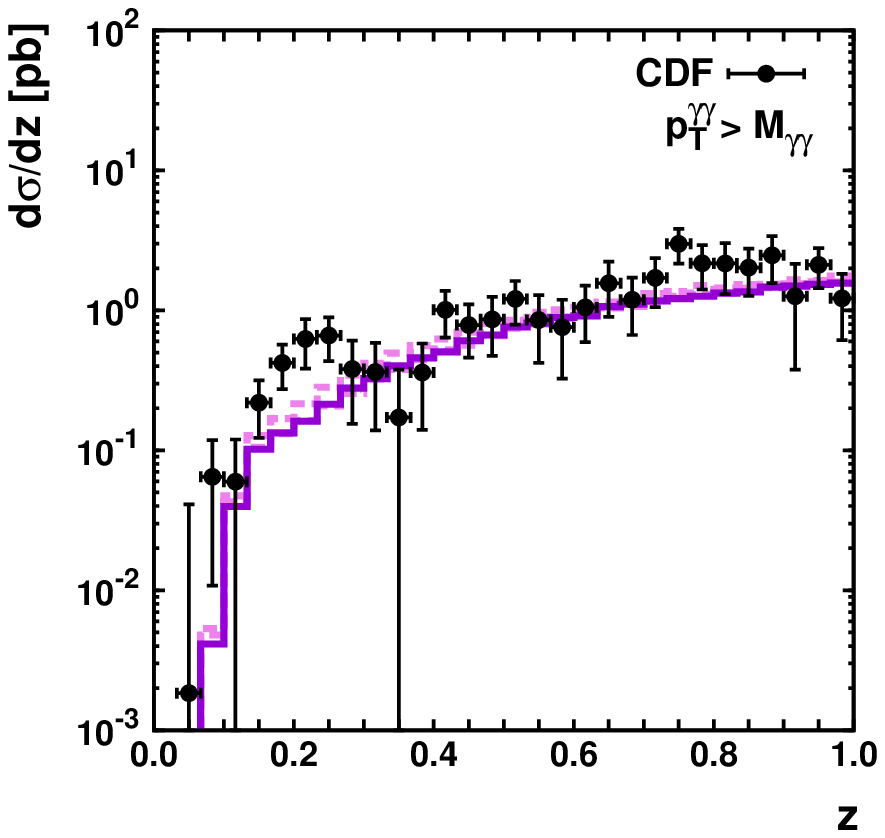, width = 8.1cm}
\caption{The differential cross section of  
prompt photon pair production in $p\bar p$ collisions
at the Tevatron as a function of diphoton rapidity $y_{\gamma \gamma}$
and variable $z = p_T^{\gamma 2}/p_T^{\gamma 1}$.
Two kinematic cases are shown separately:
differential cross sections for $M_{\gamma \gamma} > p_T^{\gamma \gamma}$ and
$M_{\gamma \gamma} < p_T^{\gamma \gamma}$.
Notation of all histograms is the same as in Fig.~7.
The experimental data are from CDF\cite{2}.}
\label{fig9}
\end{center}
\end{figure}

\begin{figure}
\begin{center}
\epsfig{figure=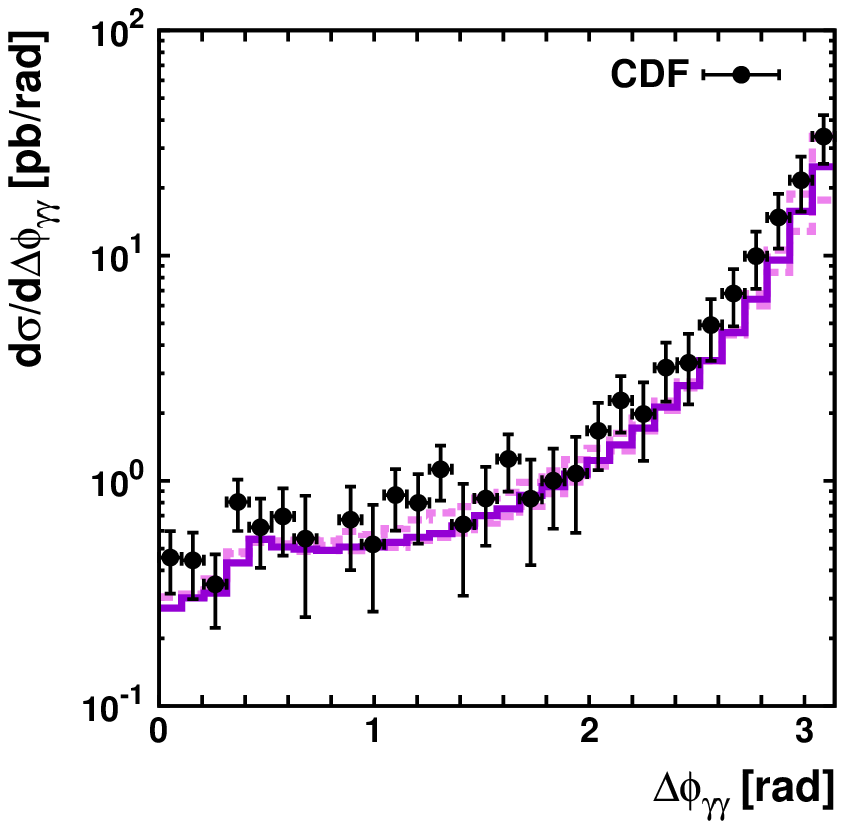, width = 8.1cm}
\epsfig{figure=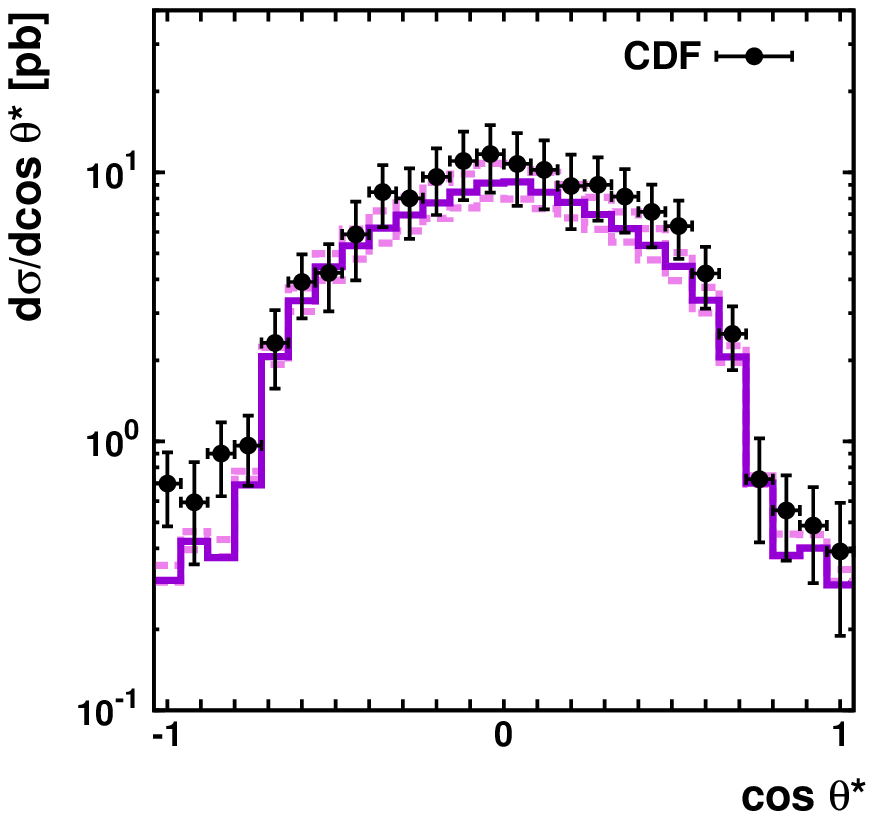, width = 8.1cm}
\epsfig{figure=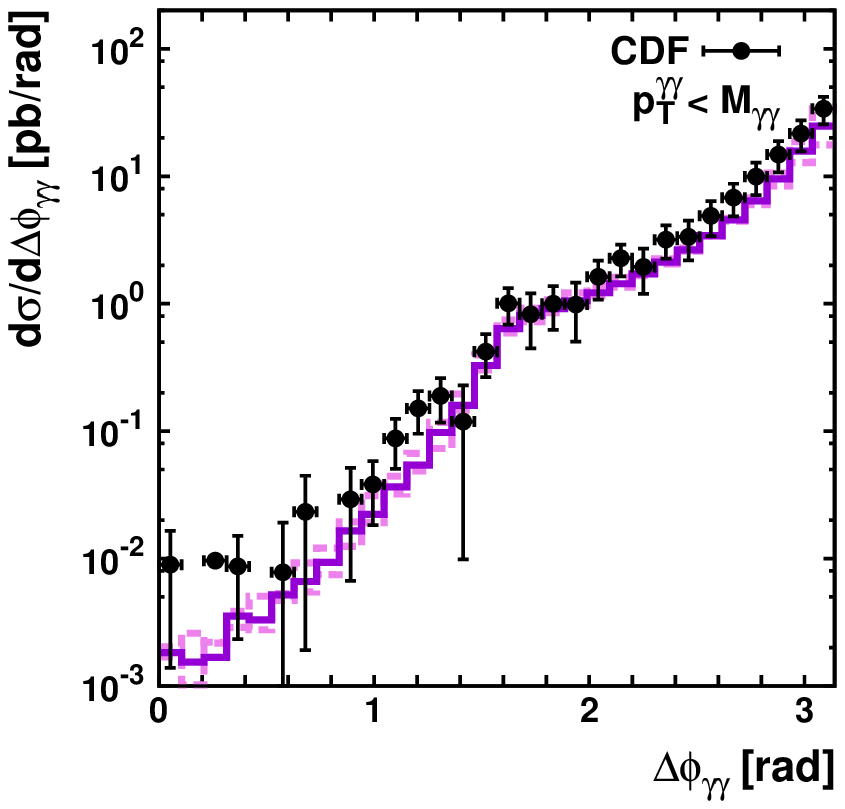, width = 8.1cm}
\epsfig{figure=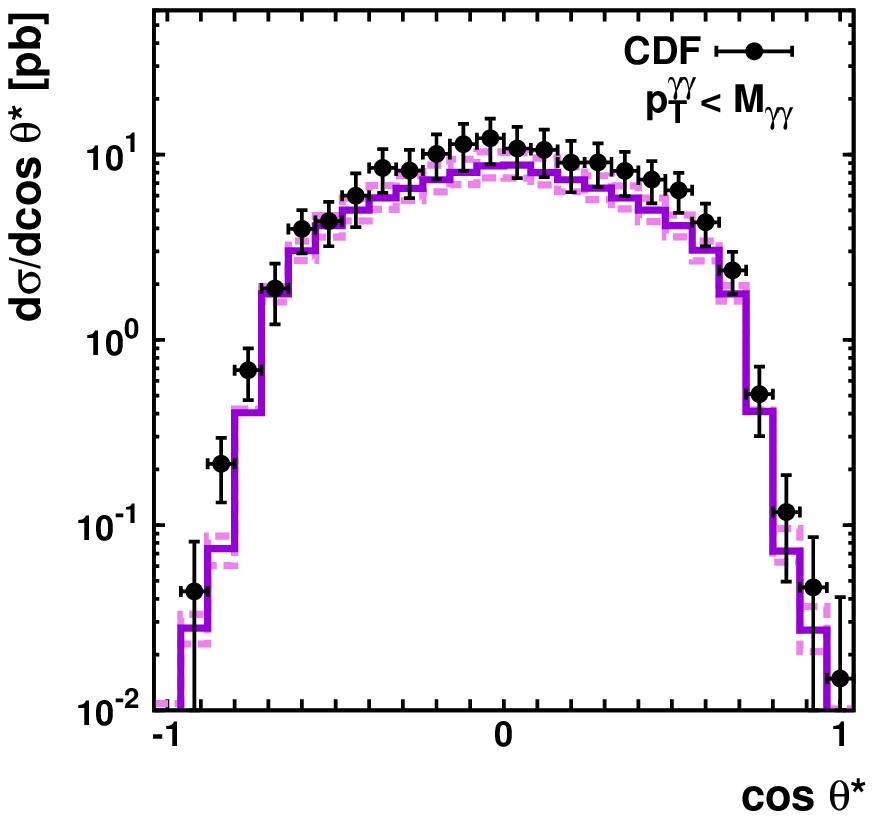, width = 8.1cm}
\epsfig{figure=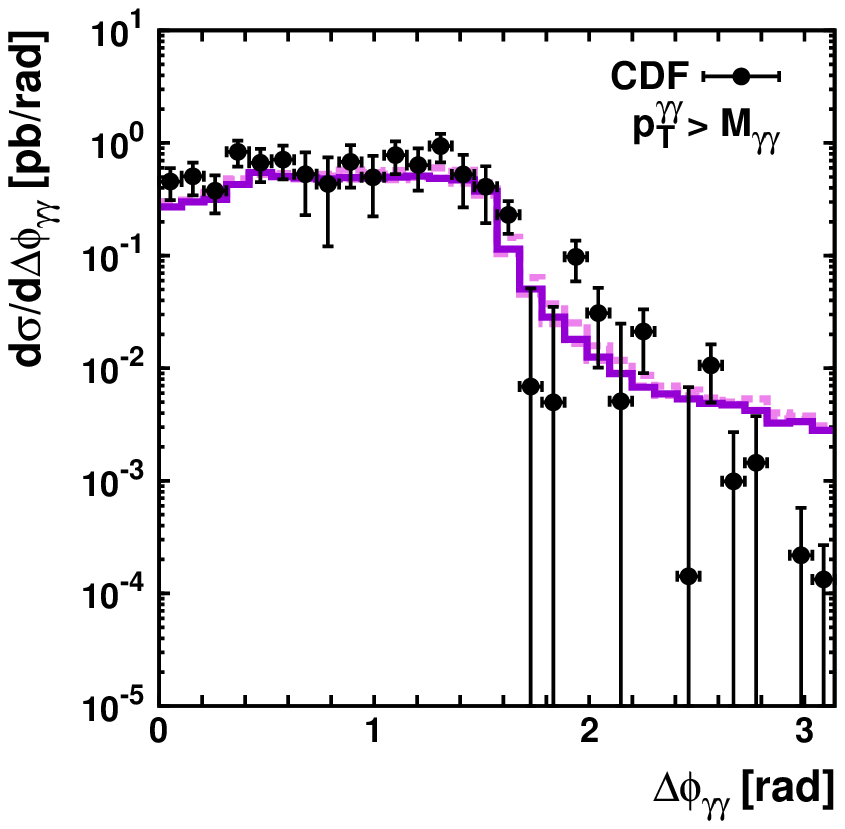, width = 8.1cm}
\epsfig{figure=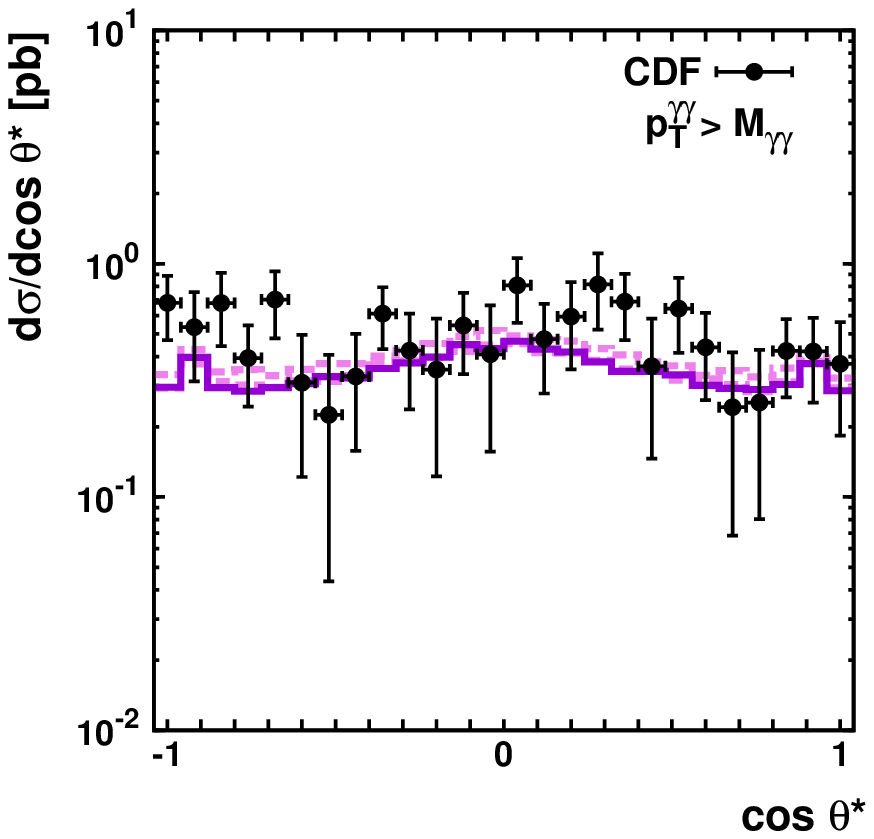, width = 8.1cm}
\caption{The differential cross section of  
prompt photon pair production in $p\bar p$ collisions
at the Tevatron as a function of $\Delta \phi_{\gamma \gamma}$
and $\cos \theta^*$.
Two kinematic cases are shown separately:
differential cross sections for $M_{\gamma \gamma} > p_T^{\gamma \gamma}$ and
$M_{\gamma \gamma} < p_T^{\gamma \gamma}$.
Notation of all histograms is the same as in Fig.~7.
The experimental data are from CDF\cite{2}.}
\label{fig10}
\end{center}
\end{figure}

\begin{figure}
\begin{center}
\epsfig{figure=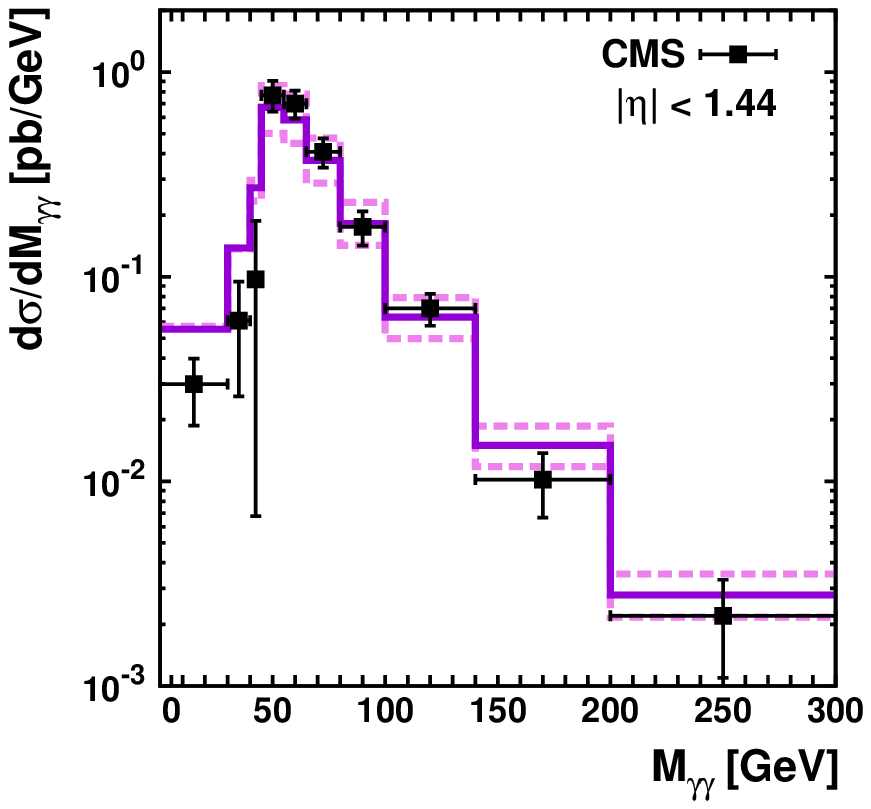, width = 8.1cm}
\epsfig{figure=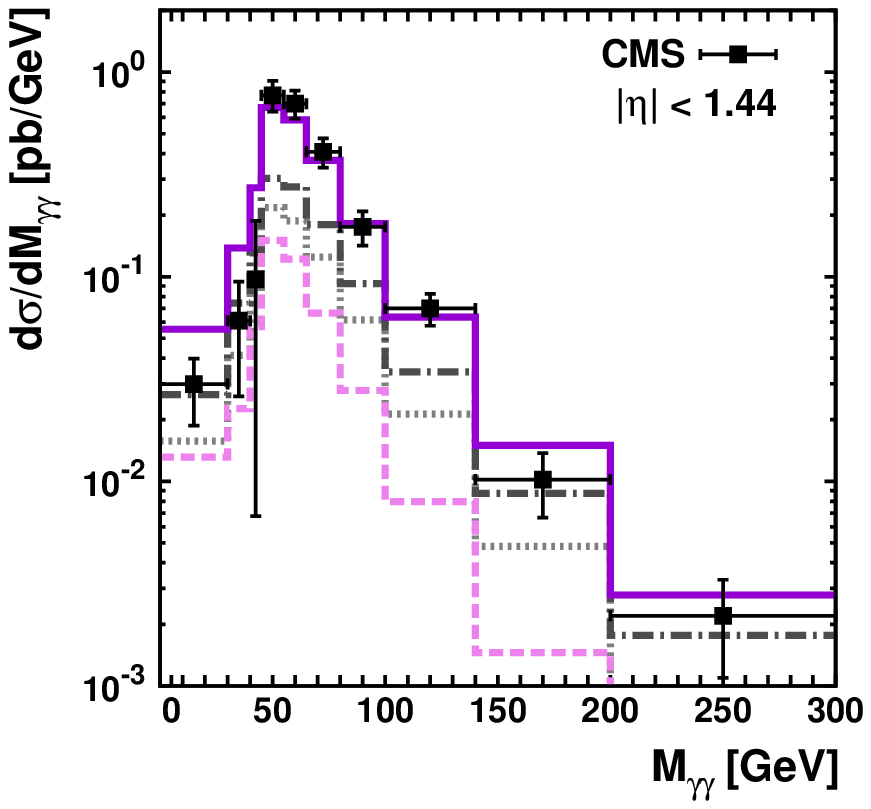, width = 8.1cm}
\epsfig{figure=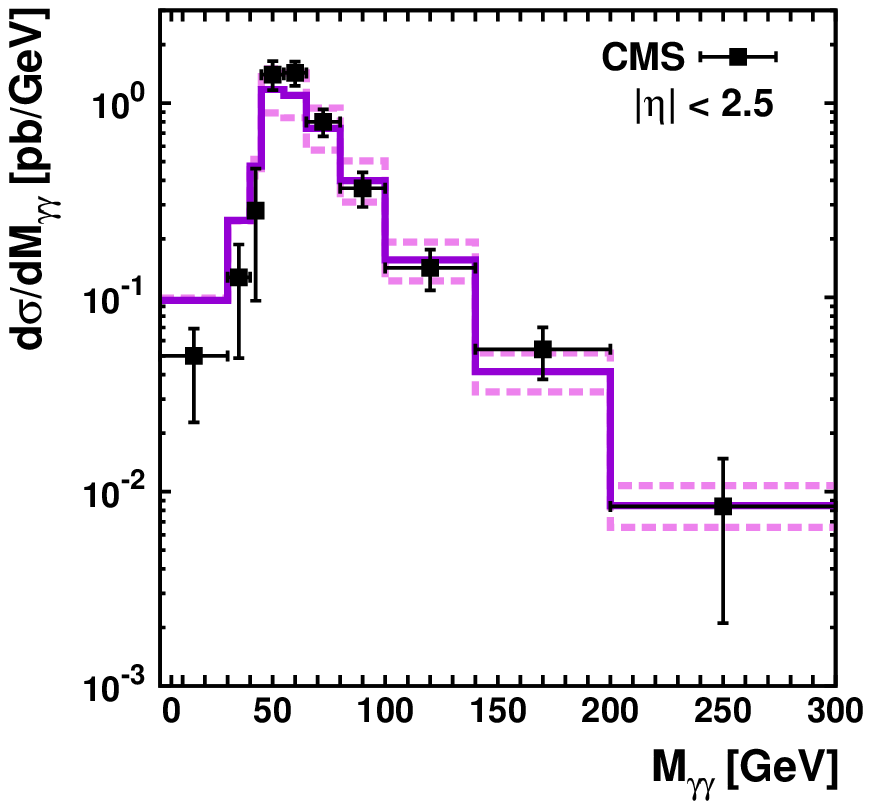, width = 8.1cm}
\epsfig{figure=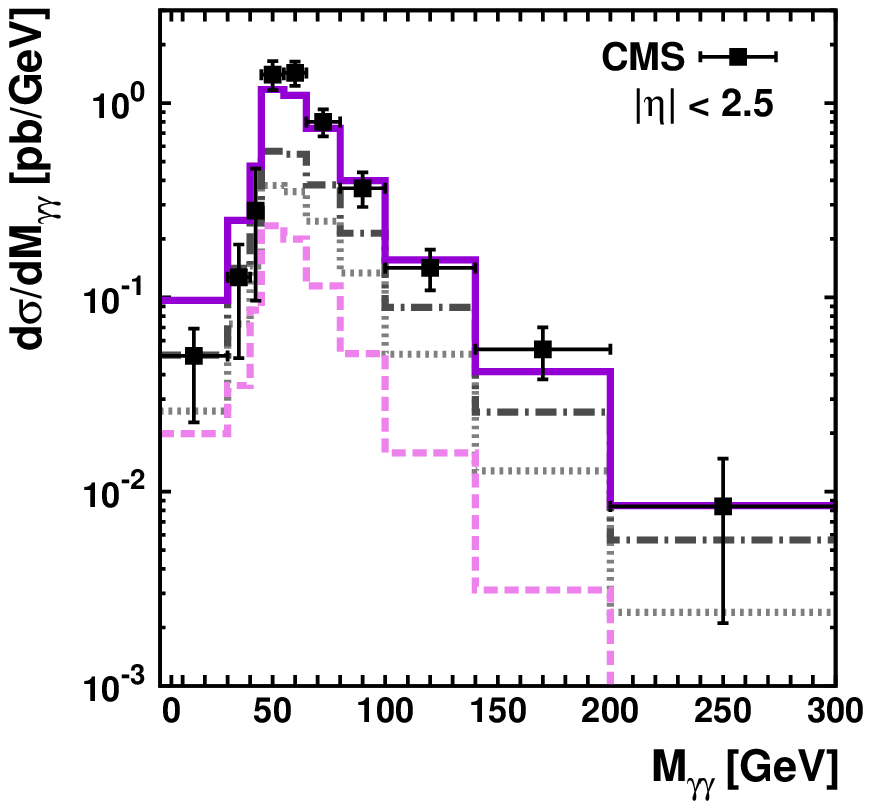, width = 8.1cm}
\caption{The differential cross section of  
prompt photon pair production in $p p$ collisions
at the LHC as a function of diphoton invariant mass.
Left panel: the solid histograms correspond to the results obtained with the KMR parton densities
at the default scale, and the upper and lower dashed histograms correspond to standard scale variations, as it is
described in the text. Right panel: the different contributions to the 
diphoton cross sections calculated at the default scale. 
The dashed, dash-dotted and dotted histograms correspond to the
gluon-gluon fusion, quark-antiquark annihilation and quark-gluon scattering subprocesses.
The solid histograms represent the sum of these contributions.
The experimental data are from CMS\cite{4}.}
\label{fig11}
\end{center}
\end{figure}

\begin{figure}
\begin{center}
\epsfig{figure=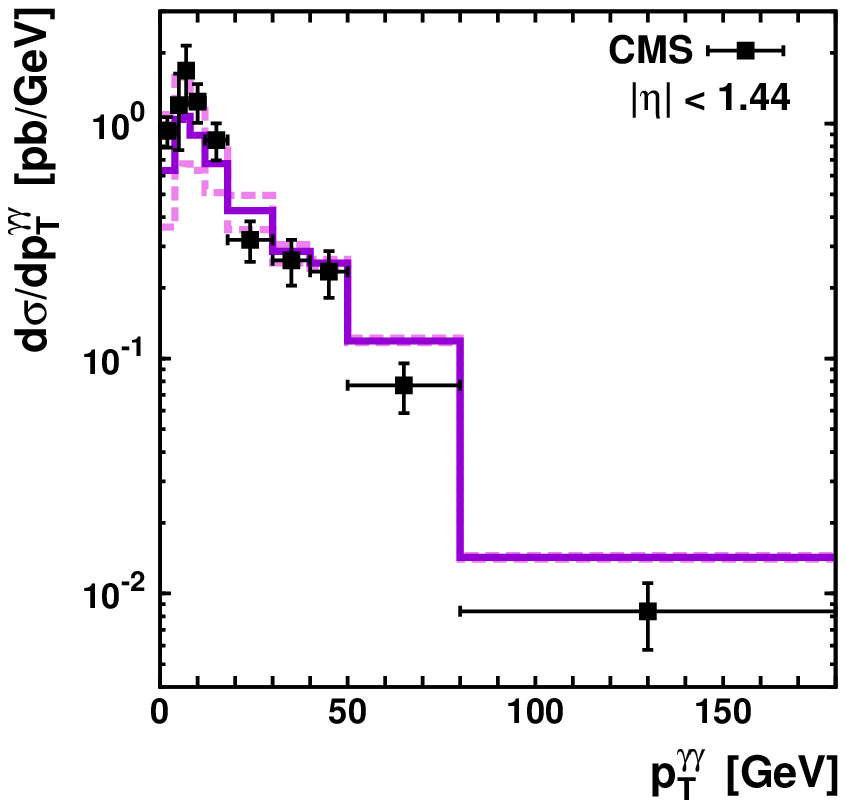, width = 8.1cm}
\epsfig{figure=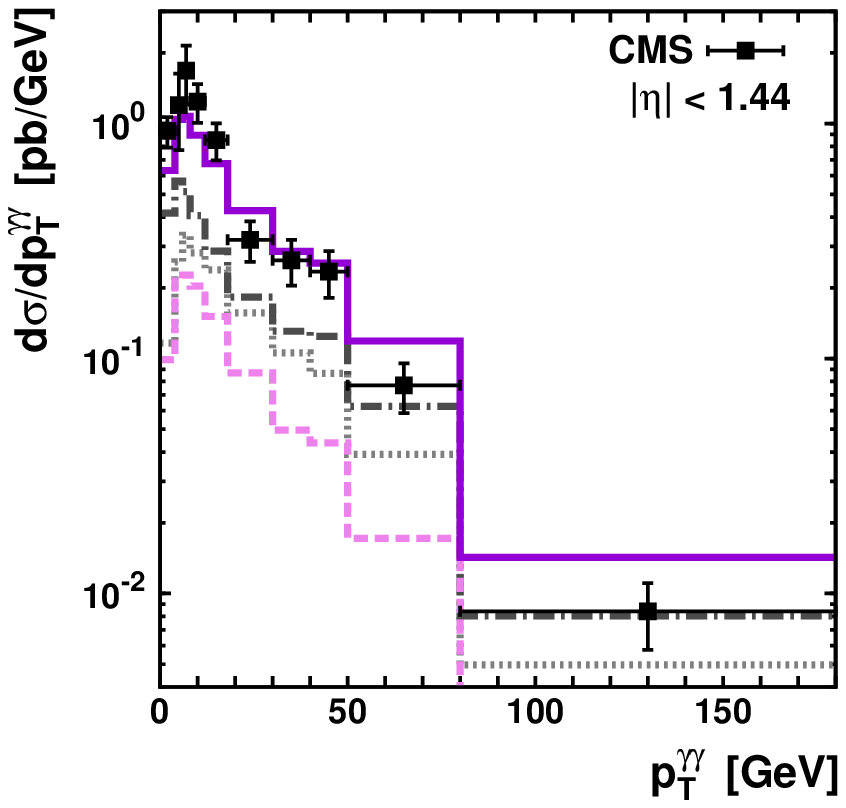, width = 8.1cm}
\epsfig{figure=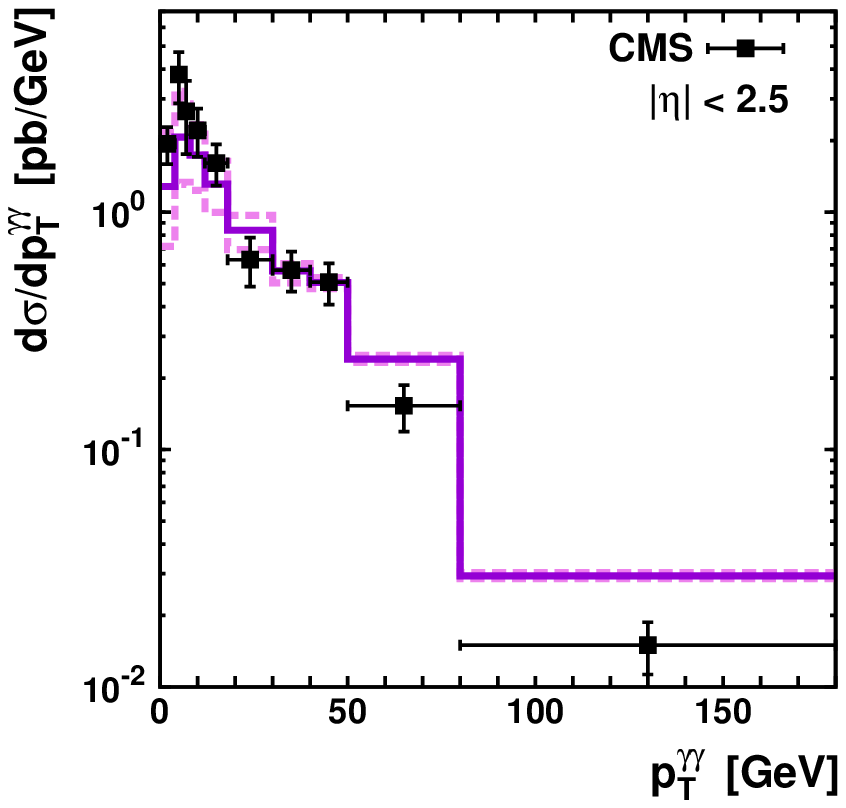, width = 8.1cm}
\epsfig{figure=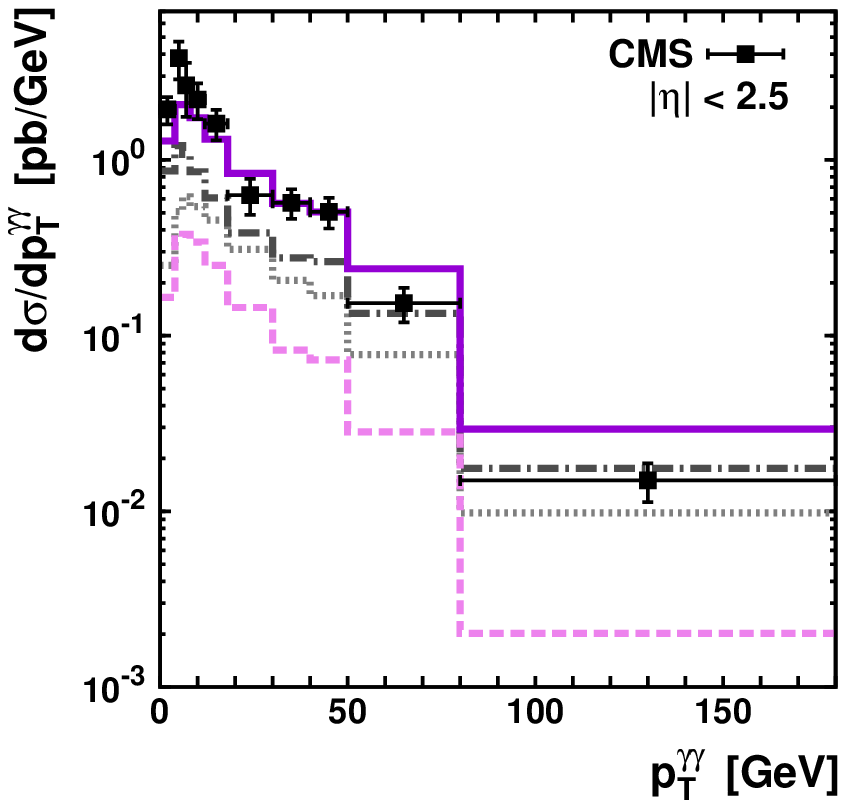, width = 8.1cm}
\caption{The differential cross section of  
prompt photon pair production in $p p$ collisions
at the LHC as a function of diphoton transverse momentum.
Notation of all histograms is the same as in Fig.~11.
The experimental data are from CMS\cite{4}.}
\label{fig12}
\end{center}
\end{figure}

\begin{figure}
\begin{center}
\epsfig{figure=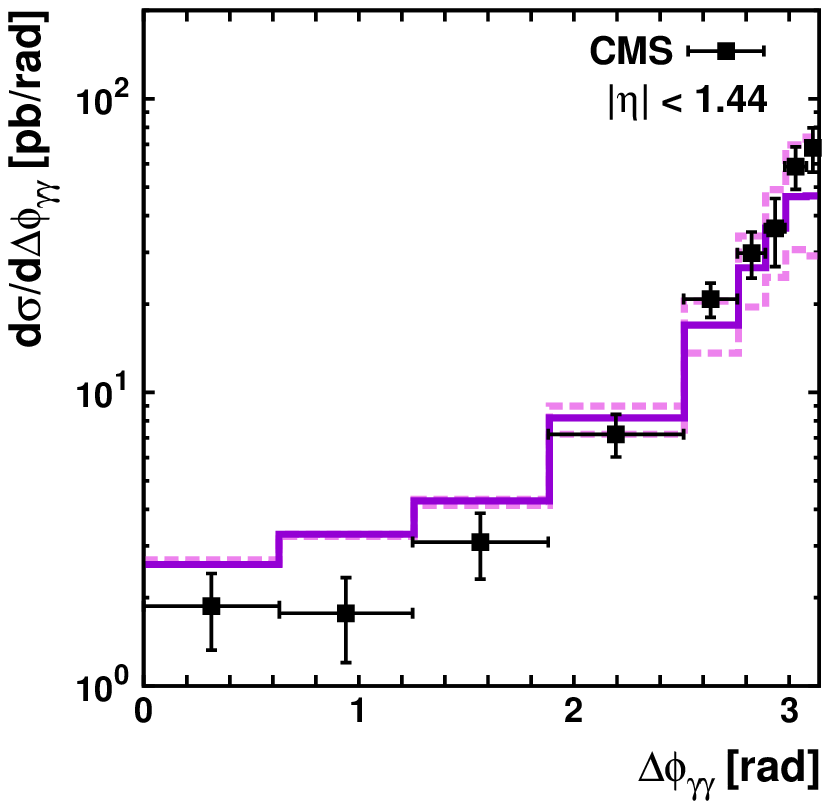, width = 8.1cm}
\epsfig{figure=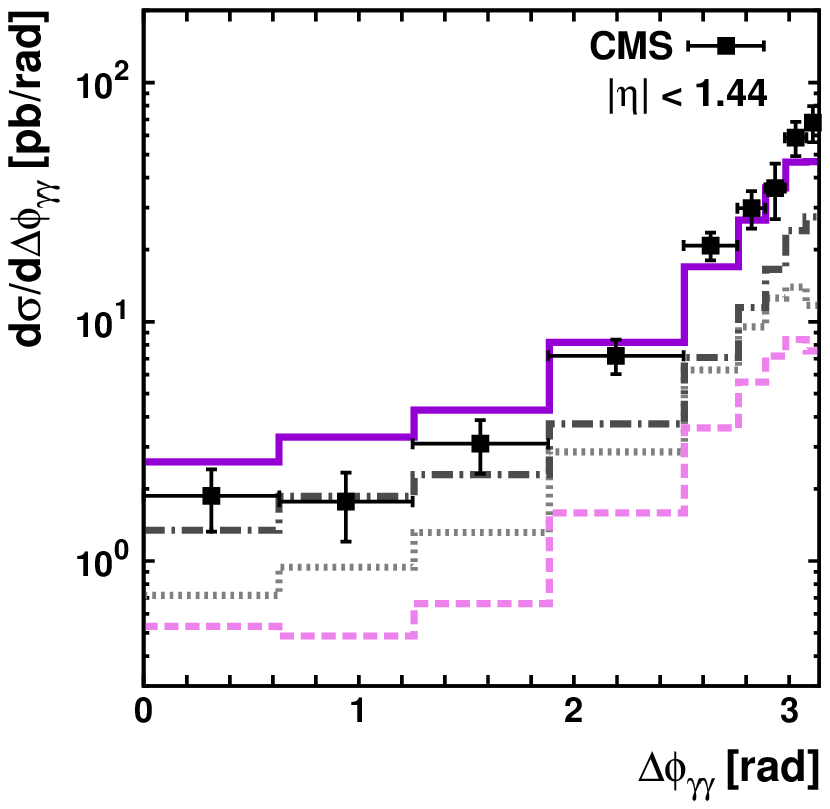, width = 8.1cm}
\epsfig{figure=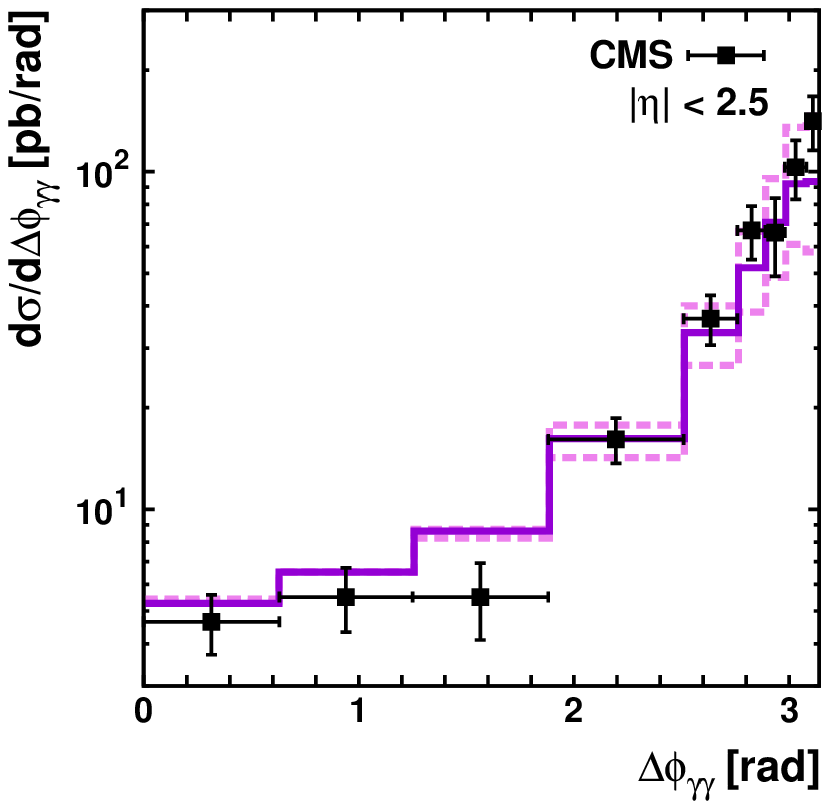, width = 8.1cm}
\epsfig{figure=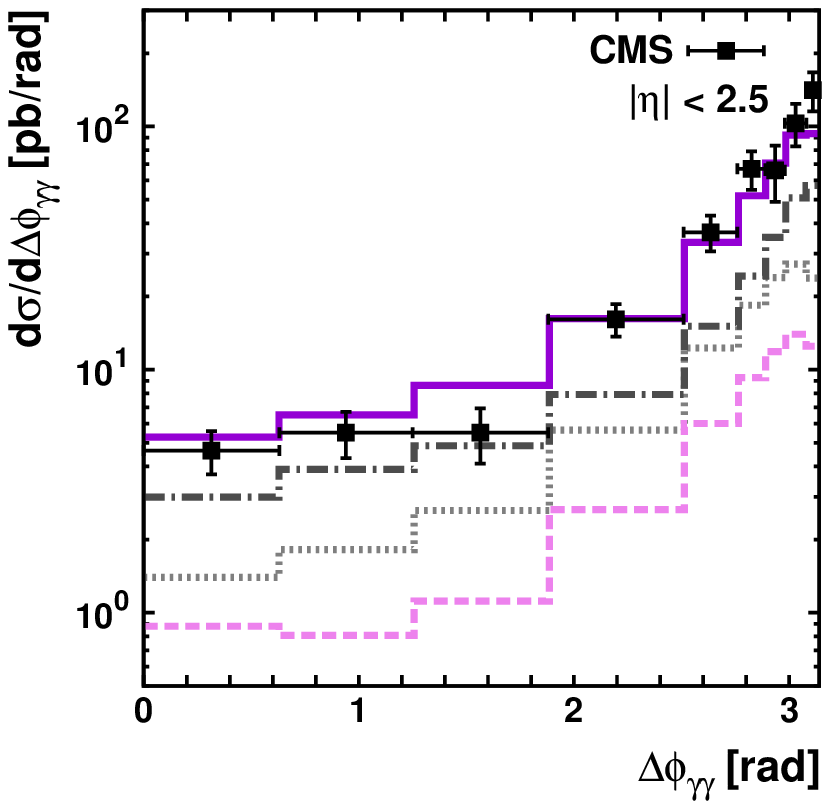, width = 8.1cm}
\caption{The differential cross section of  
prompt photon pair production in $p p$ collisions
at the LHC as a function of $\Delta \phi_{\gamma \gamma}$.
Notation of all histograms is the same as in Fig.~11.
The experimental data are from CMS\cite{4}.}
\label{fig13}
\end{center}
\end{figure}

\begin{figure}
\begin{center}
\epsfig{figure=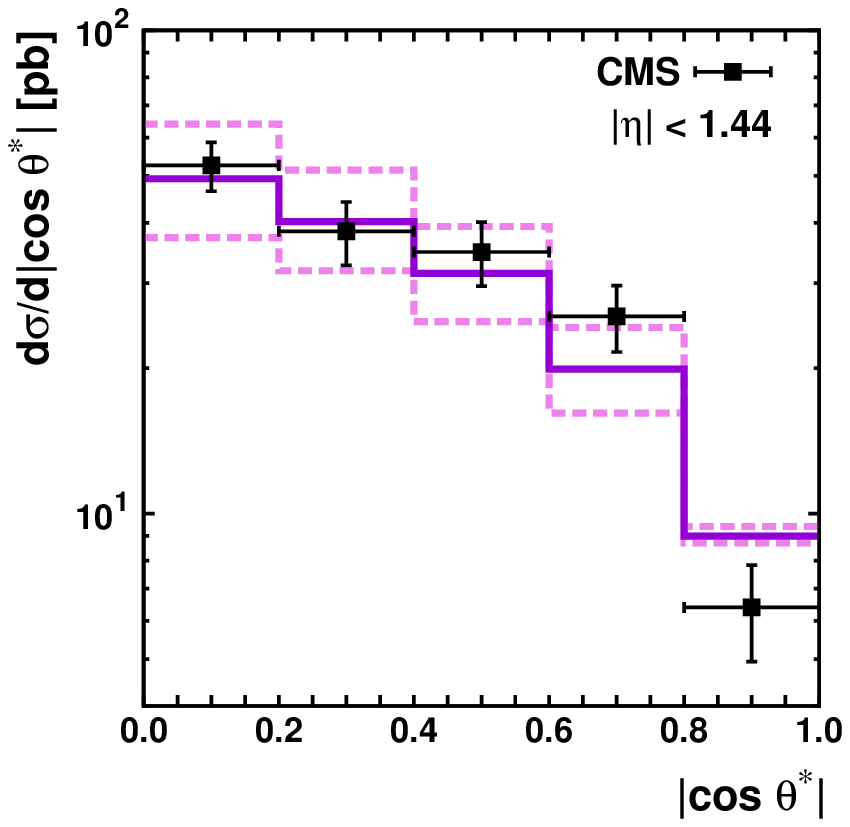, width = 8.1cm}
\epsfig{figure=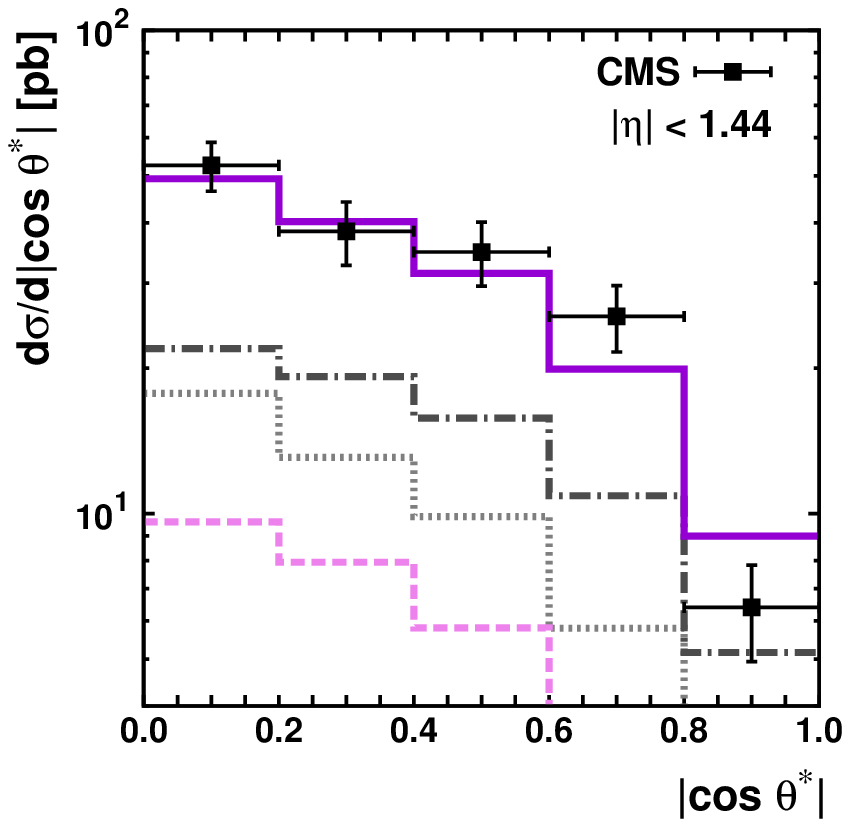, width = 8.1cm}
\epsfig{figure=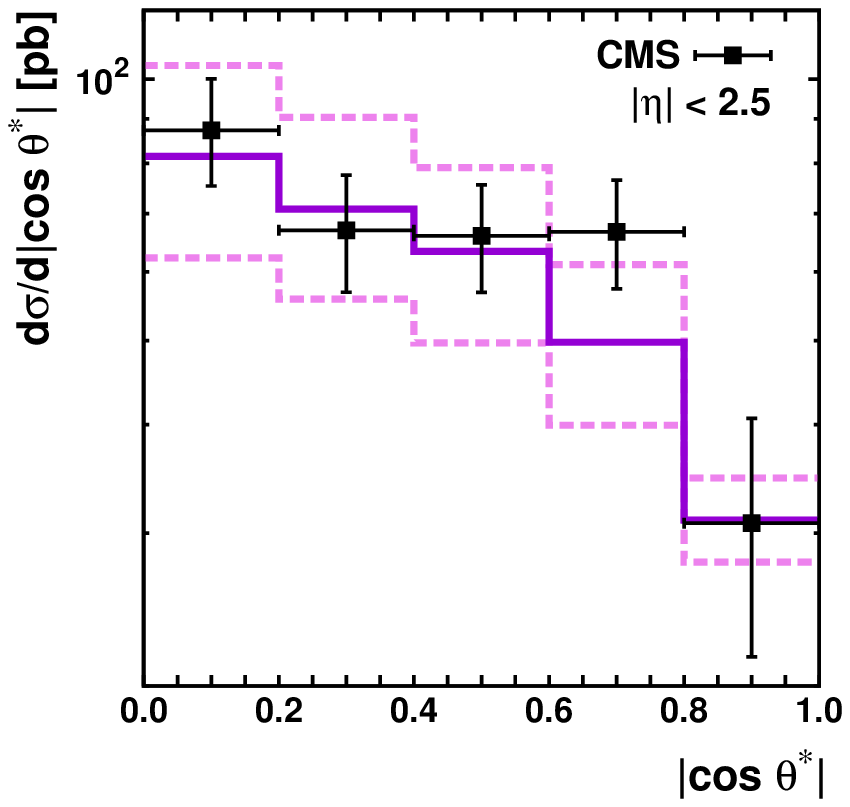, width = 8.1cm}
\epsfig{figure=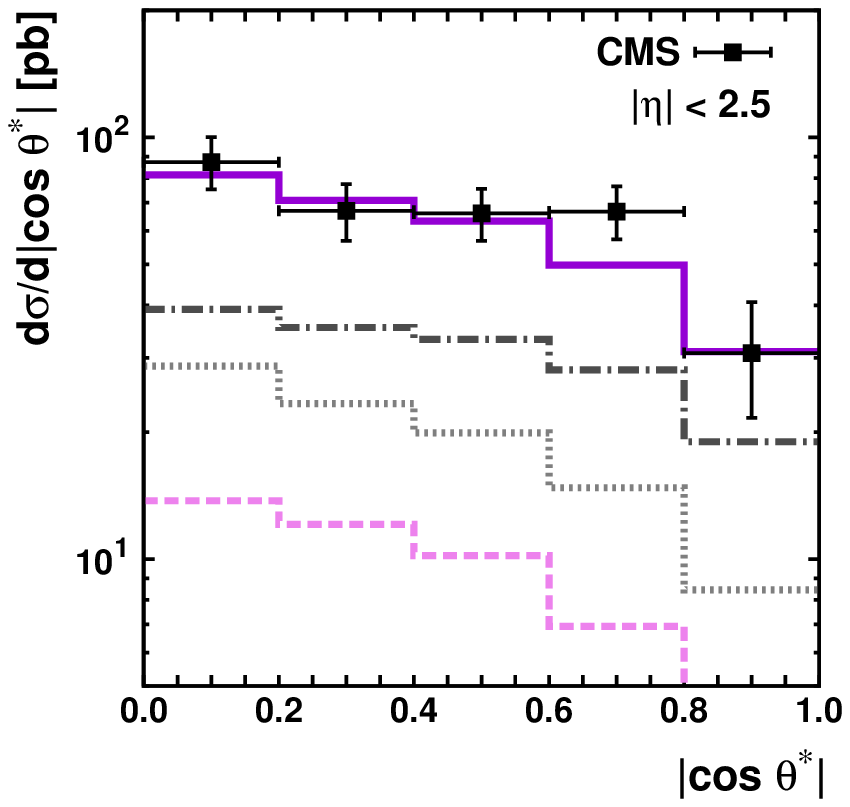, width = 8.1cm}
\caption{The differential cross section of  
prompt photon pair production in $p p$ collisions
at the LHC as a function of $|\cos \theta^*|$.
Notation of all histograms is the same as in Fig.~11.
The experimental data are from CMS\cite{4}.}
\label{fig14}
\end{center}
\end{figure}

\begin{figure}
\begin{center}
\epsfig{figure=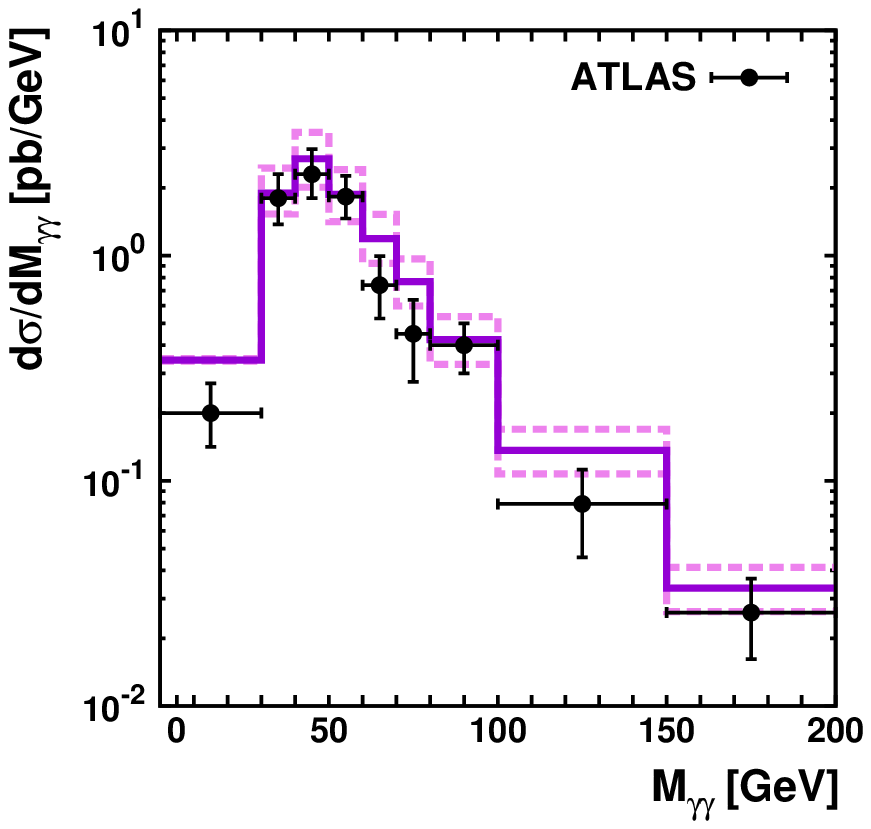, width = 8.1cm}
\epsfig{figure=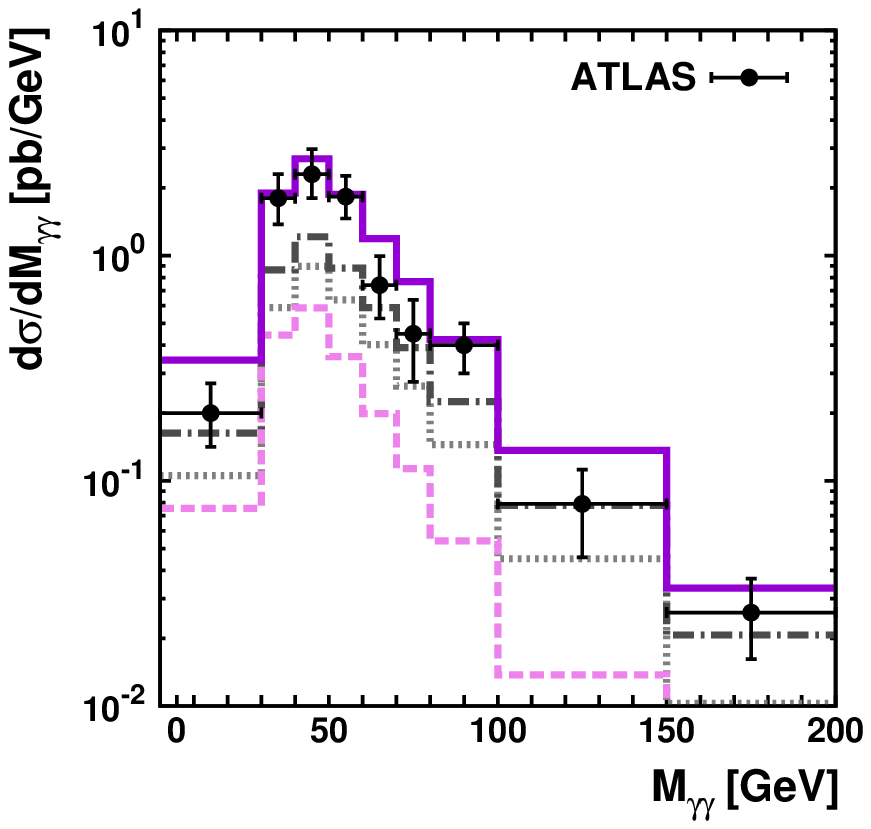, width = 8.1cm}
\epsfig{figure=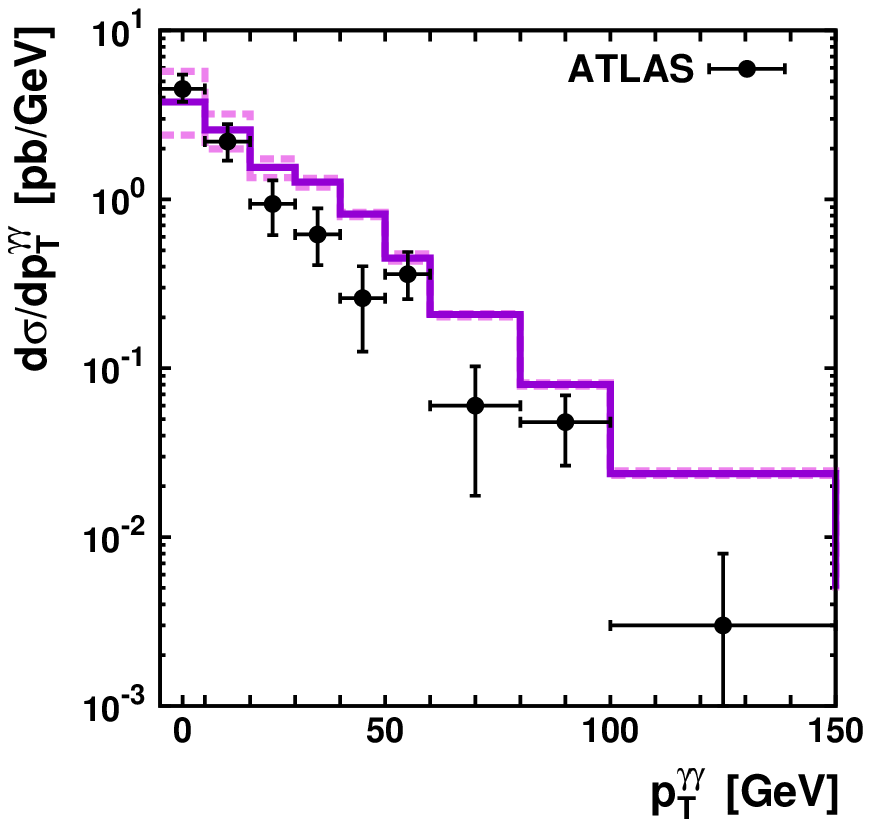, width = 8.1cm}
\epsfig{figure=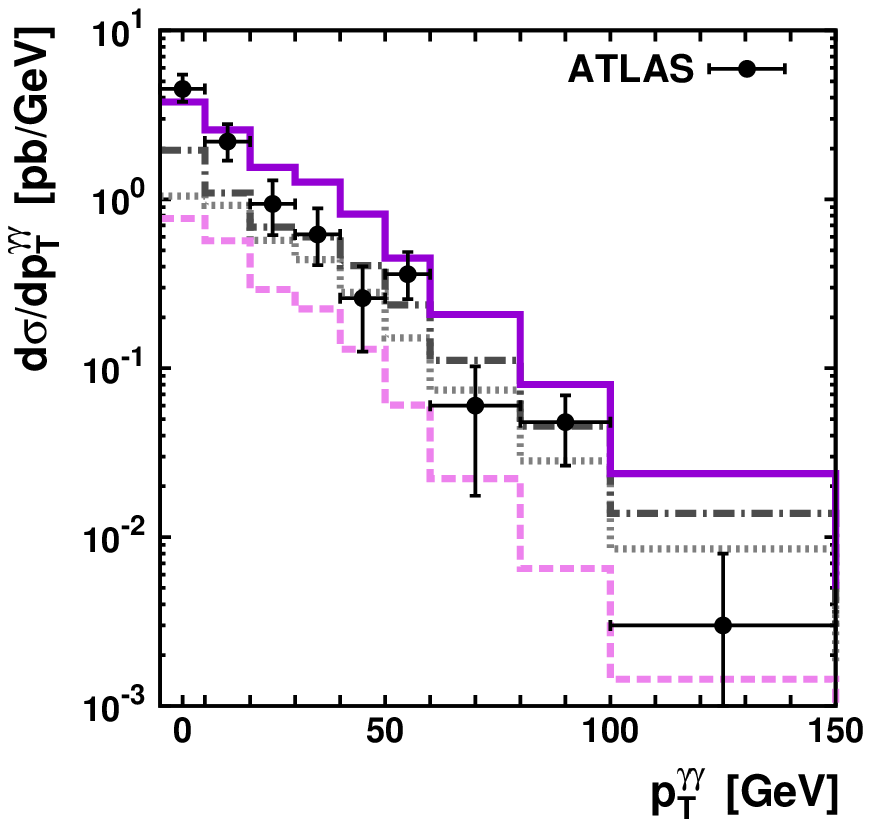, width = 8.1cm}
\epsfig{figure=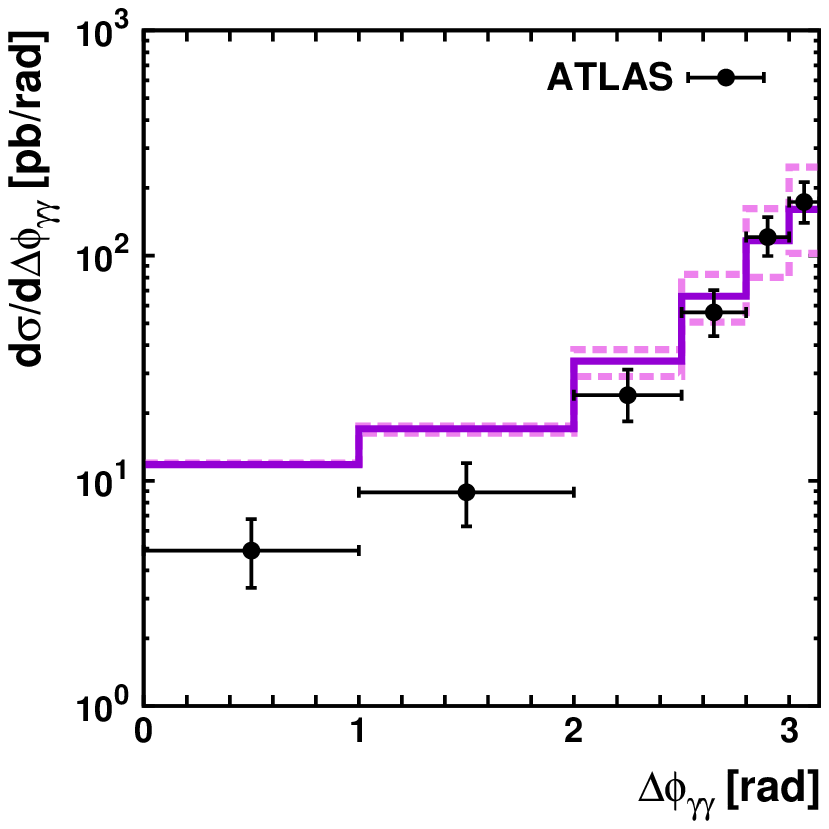, width = 8.1cm}
\epsfig{figure=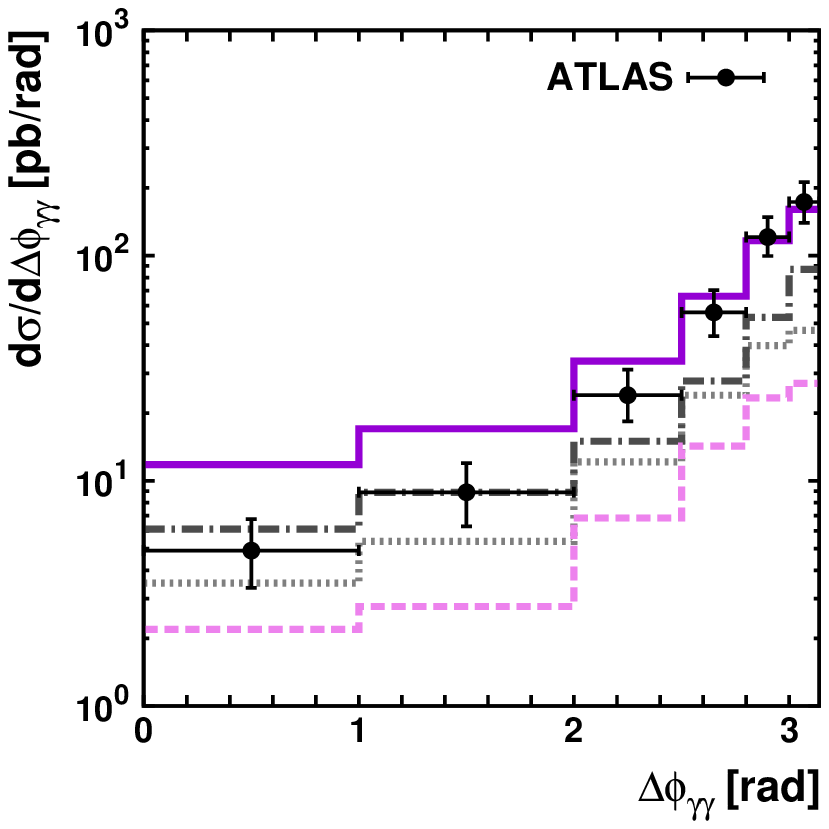, width = 8.1cm}
\caption{The differential cross section of  
prompt photon pair production in $p p$ collisions
at the LHC.
Notation of all histograms is the same as in Fig.~11.
The experimental data are from ATLAS\cite{5}.}
\label{fig15}
\end{center}
\end{figure}


\begin{thebibliography}{36}

\bibitem{1} D.~Acosta {\sl et al.} (CDF Collaboration), Phys. Rev. Lett. {\bf 95}, 022003 (2005). 
\bibitem{2} T.~Aaltonen {\sl et al.} (CDF Collaboration), Phys. Rev. D {\bf 84}, 052006 (2011). 
\bibitem{3} V.~Abazov {\sl et al.} (D$\emptyset$ Collaboration), Phys. Lett. B {\bf 690}, 108 (2010). 
\bibitem{4} CMS Collaboration, JHEP {\bf 01}, 133 (2012).
\bibitem{5} ATLAS Collaboration, Phys. Rev. D {\bf 85}, 012003 (2012).
\bibitem{6} S.~Mrenna and J.~Willis, Phys. Rev. D {\bf 63}, 015006 (2001).
\bibitem{7} G.F.~Giudice and R.~Rattazzi, Phys. Rep. {\bf 322}, 419 (1999).
\bibitem{8} T.~Appelquist, H.~Cheng, and B.~Dobrescu, Phys. Rev. D {\bf 64}, 035002 (2001).
\bibitem{9} ATLAS Collaboration, Phys. Rev. Lett. {\bf 106}, 121803 (2011).
\bibitem{10} L.~Randall and R.~Sundrum, Phys. Rev. Lett. {\bf 83}, 3370 (1999).
\bibitem{11} K.~Koller, T.F.~Walsh, and P.M.~Zerwas, Z. Phys. C {\bf 2}, 197 (1979).
\bibitem{12} T.~Binoth, J.P.~Guillet, E.~Pilon, and M.~Werlen, Eur. Phys. J. C {\bf 16}, 311 (2000);\\
 T.~Binoth, J.P.~Guillet, E.~Pilon, and M.~Werlen, Phys. Rev. D {\bf 63}, 114016 (2001).
\bibitem{13} C.~Balazs, E.L.~Berger, P.~Nadolsky and C.-P.~Yuan, Phys. Rev. D {\bf 76}, 013009 (2007).
\bibitem{14} L.V.~Gribov, E.M.~Levin, and M.G.~Ryskin, Phys. Rep. {\bf 100}, 1 (1983);\\
  E.M.~Levin, M.G.~Ryskin, Yu.M.~Shabelsky and A.G.~Shuvaev, Sov. J. Nucl. Phys. {\bf 53}, 657 (1991).
\bibitem{15}  S.~Catani, M.~Ciafoloni and F.~Hautmann, Nucl. Phys. B {\bf 366}, 135 (1991);\\
  J.C.~Collins and R.K.~Ellis, Nucl. Phys. B {\bf 360}, 3 (1991).
\bibitem{16} E.A.~Kuraev, L.N.~Lipatov and V.S.~Fadin, Sov. Phys. JETP {\bf 44}, 443 (1976);\\
  E.A.~Kuraev, L.N.~Lipatov and V.S.~Fadin, Sov. Phys. JETP {\bf 45}, 199 (1977);\\
  I.I.~Balitsky and L.N.~Lipatov, Sov. J. Nucl. Phys. {\bf 28}, 822 (1978).
\bibitem{17} M.~Ciafaloni, Nucl. Phys. B {\bf 296}, 49 (1988);\\
  S.~Catani, F.~Fiorani and G.~Marchesini, Phys. Lett. B {\bf 234}, 339 (1990);\\
  S.~Catani, F.~Fiorani and G.~Marchesini, Nucl. Phys. B {\bf 336}, 18 (1990);\\
  G.~Marchesini, Nucl. Phys. B {\bf 445}, 49 (1995).
\bibitem{18} B.~Andersson {\sl et al.} (Small-$x$ Collaboration), Eur. Phys. J. C {\bf 25}, 77 (2002);\\
  J.~Andersen {\sl et al.} (Small-$x$ Collaboration), Eur. Phys. J. C {\bf 35}, 67 (2004);\\
  J.~Andersen {\sl et al.} (Small-$x$ Collaboration), Eur. Phys. J. C {\bf 48}, 53 (2006).
\bibitem{19} A.~Gawron and J.~Kwiecinski, Phys. Rev. D {\bf 70}, 014003 (2004).
\bibitem{20} S.P.~Baranov, A.V.~Lipatov and N.P.~Zotov, Phys. Rev. D {\bf 81}, 094034 (2010);\\
  A.V.~Lipatov and N.P.~Zotov, Phys. Rev. D {\bf 81}, 094027 (2010); Phys. Rev. D {\bf 72}, 054002 (2005).
\bibitem{21} A.V.~Lipatov and N.P.~Zotov, J. Phys. G {\bf 34}, 219 (2007).
\bibitem{22} S.P.~Baranov, A.V.~Lipatov and N.P.~Zotov, Phys. Rev. D {\bf 77}, 074024 (2008).
\bibitem{23} A.V.~Lipatov, M.A.~Malyshev and N.P.~Zotov, Phys. Lett. B {\bf 699}, 93 (2011).
\bibitem{24} A.V.~Lipatov, M.A.~Malyshev and N.P.~Zotov, JHEP {\bf 1112}, 117 (2011).
\bibitem{25} A.V.~Lipatov, M.A.~Malyshev and N.P.~Zotov, JHEP {\bf 1205}, 104 (2012).
\bibitem{26} M.A.~Kimber, A.D.~Martin and M.G.~Ryskin, Phys. Rev. D {\bf 63}, 114027 (2001);\\
  G.~Watt, A.D.~Martin and M.G.~Ryskin, Eur. Phys. J. C {\bf 31}, 73 (2003).
\bibitem{27} R.E.~Prange, Phys. Rev. {\bf 110}, 240 (1958);\\
  S.P.~Baranov, Phys. Atom. Nucl. {\bf 60}, 1322 (1997).
\bibitem{28} J.A.M.~Vermaseren, ”Symbolic Manipulation with FORM”, published by Computer
Algebra Nederland, Kruislaan 413, 1098, SJ Amsterdaam, 1991; ISBN 90-74116-01-9.
\bibitem{29} E.L.~Berger, E.~Braaten, and R.D.~Field, Nucl. Phys. B  {\bf 239}, 52 (1984).
\bibitem{30} A.D.~Martin, W.J.~Stirling, R.S.~Thorne and G.~Watt, Eur. Phys. J. C {\bf 63}, 189 (2009).
\bibitem{31} M.~Fontannaz, J.Ph.~Guillet and G.~Heinrich, Eur. Phys. J. C {\bf 21}, 303 (2001).
\bibitem{32} G.P.~Lepage, J. Comput. Phys. {\bf 27}, 192 (1978).
\bibitem{33} S.P.~Baranov, A.V.~Lipatov and N.P.~Zotov, Yad. Fiz. {\bf 67}, 856 (2004).
\bibitem{34} V.A.~Saleev, Phys. Rev. D {\bf 80}, 114016 (2009).
\bibitem{35} S.P.~Baranov, A.V.~Lipatov and N.P.~Zotov, Phys. Rev. D {\bf 85}, 014034 (2012);\\
  S.P.~Baranov, A.V.~Lipatov and N.P.~Zotov, Eur. Phys. J. C {\bf 71}, 1631 (2011).
\bibitem{36} S.P.~Baranov and N.P.~Zotov, JETP Lett. {\bf 88}, 711 (2008).

\end{thebibliography}
\end{document}